\documentclass[11pt, onecolumn]{article}
\usepackage[margin = 1in]{geometry}

\usepackage{authblk}
\usepackage{amssymb}
\usepackage{booktabs}
\usepackage{amsmath}

\usepackage{tikz}
\usepackage{pgf}
\usetikzlibrary{patterns,arrows,decorations.pathreplacing}

\usepackage{amsthm}
\usepackage{url}
\usepackage{algorithm}
\usepackage{algorithmic}
\usepackage{trivfloat}
\trivfloat{program}

\usepackage{float}
\usepackage{epstopdf}
\usepackage{verbatim}
\usepackage{empheq}
\usepackage{hyperref}
\usepackage{color}
\usepackage{bigfoot}
\usepackage{slashbox} 
\usepackage[font=footnotesize]{caption}

\usepackage[us, 12hr]{datetime}

\usepackage{multirow}
\usepackage{setspace}
\newcommand{\tp}{\mathsf{T}}				

\newcommand{\real} 	  {\mathbb{R}}

\newcommand{\half}{\frac{1}{2}}

\newcommand{\ds}{\displaystyle}	

\newcommand{\opt}{^{\mathsf{opt}}}
\newcommand{\major}{^{\mathsf{M}}}

\newcommand{\lam} {\lambda}	
\newcommand{\eps} {\epsilon}	

\newcommand{\phie} {\phi_{\epsilon}}

\providecommand{\norm}[1]{\lVert#1\rVert}
\providecommand{\abs}[1]{\left\vert #1 \right\vert}

\newtheorem{proposition}{Proposition}

\providecommand{\tabref}[1]{Table~\ref{#1}}

\providecommand{\figref}[1]{Figure~\ref{#1}}
\providecommand{\appref}[1]{Appendix~\ref{#1}}

\providecommand{\eqnref}[1]{\eqref{#1}}

\newcommand{\Figurescale}		{0.625}				
\newcommand{\figurescale}		{0.470}				

\allowdisplaybreaks

\title{Repetitive Transients Extraction Algorithm \\for Detecting Bearing Faults%
\footnote{Preprint}}

\author[1]{Wangpeng He}
\author[2]{Yin Ding\thanks{email: yd372@nyu.edu } }
\author[3 4]{Yanyang Zi}
\author[2]{Ivan W. Selesnick}

\affil[1]{School of Aerospace Science and Technology, Xidian University, Xi'an, China} 
\affil[2]{Tandon School of Engineering, New York University, 6 Metrotech Center, Brooklyn, NY, USA}
\affil[3]{State Key Laboratory for Manufacturing and Systems Engineering, Xi'an~Jiaotong University, Xi'an, China}
\affil[4]{School of Mechanical Engineering, Xi'an~Jiaotong University, Xi'an, China}

\begin{document}

\maketitle
\begin{abstract}

This paper addresses the problem of noise reduction with simultaneous components extraction in vibration signals for faults diagnosis of bearing.
The observed vibration signal is modeled as a summation of two components contaminated by noise,
and each component composes of repetitive transients.
To extract the two components simultaneously,
an approach by solving an optimization problem is proposed in this paper.
The problem adopts convex sparsity-based regularization scheme for decomposition,
and non-convex regularization is used to further promote the sparsity but preserving the global convexity.
A synthetic example is presented to illustrate the performance of the proposed approach for repetitive feature extraction.
The performance and effectiveness of the proposed method are further demonstrated by applying to compound faults and single fault diagnosis of a locomotive bearing.
The results show the proposed approach can effectively extract the features of outer and inner race defects.
\end{abstract}


\newpage
\section{Introduction}

Rolling bearings are one of the most prevalent components in rotating machines and reciprocating machines \cite{randall2011rolling}.
Vibration-based fault detection has become the preferred technique for bearing fault diagnosis \cite{yan2014wavelets, cui2016vibration}.
Bearing vibrations are usually modeled as cyclostationary signals (or group-sparse periodic) \cite{boustany2005subspace, boustany2008blind, fd_Antoni_mssp_2009}.
When a localized defect occurs on the bearing, repetitive transients will be generated due to the passing of rollers over the defect 
\cite{he2014automatic, smith2015rolling}.
These transients have repetitive structure and are usually submerged in background noise.
Many denoising methods have been introduced to extract fault features for the purpose of detecting faults in machines, 
such as wavelet transform  \cite{yan2010harmonic, chen2012fault}, 
singular value decomposition (SVD) \cite{jiang2015study}, 
time-frequency analysis methods \cite{feng2013recent, he2013time}, 
empirical mode decomposition (EMD) \cite{lei2013review},
methods for blind extraction of a cyclostationary signal and spectral kurtosis (SK) \cite{boustany2005subspace, boustany2008blind, antoni2006spectral}.
If compound faults exist, then the observed vibration signals are rather complex and it is difficult to identify each fault using traditional signal processing methods.

A number of approaches have been developed for the multiple fault or compound fault diagnosis.
Principal component analysis (PCA) based method has been used  to classify faults \cite{lei2008new, li2011virtual, malhi2004pca}.
Support vector machine (SVM) based methods have been used for multi-fault diagnosis \cite{zhang2013multi,abbasion2007rolling} as well.
Some other techniques such as neural network and independent component analysis (ICA),
have been introduced to assist fault detection and classification \cite{wang2011constrained, boukra2013statistical, bin2012early}.
Some of these methods require large data collection as a training set
and further require an off-line training phase;
and some methods rely on careful selection of features (e.g., wavelet packet sub-bands).

Adopting sparsity in the field of fault detection was initially illustrated in Ref.~\cite{fd_Yang_mssp_2005},
where basis pursuit denoising (BPD) \cite{Chen_1994_BP} was used to exploit sparse features in various domains to detect faults.
Some recent works consider other sparse representations for fault diagnosis
\cite{yan2014wavelets, fd_Cui_jsv_2014, he2013tunable, cui2016quantitative}.
The fault frequencies of potential bearing fault features
can be simply obtained using the geometry of the components in many cases
or directly obtained from the user operation manual.
Many works have considered the fault frequencies as priori information
\cite{yan2014wavelets, fd_Liang_mssp_2010, fd_Su_mssp_2010, fd_Qin_jsv_2013, He_mssp_2016}.

In this work, a method using the temporal periodicity
(namely fault characteristic or fundamental frequency) directly in the time domain is proposed
to extract fault features while simultaneously denoising.
The proposed method is based on convex optimization using non-convex regularization.
Specifically, this paper aims to address the problem of extracting compound features caused by faults in vibration signals,
where the features exhibit repetitive group sparsity.

In particular, the observed signal is modeled as
\begin{align}\label{eqn:fdmca_model}
	y = x_1 + x_2 + w,
\end{align}
where $w$ denotes additive white Gaussian noise (AWGN),
and $x_1$ and $x_2$ are both repetitively group-sparse signals with
periods $T_1$ and $T_2$ respectively.
Note that here the repetitively group-sparse signal means the useful features (group-sparse) appear repetitively. 
In other words, the features have group-sparse property and appear repetitively.
To avoid confusion, here the ``period" is referred to as the cycle for cyclostationary signal in Ref.~\cite{fd_Antoni_mssp_2007}.
In the case of compound faults detection of bearings, useful features $x_1$ and $x_2$ also satisfy the following two conditions.
\begin{enumerate}
\item
	The periods $T_1$ and $T_2$ are different.
\item
	Each period is not close to an integer multiple of the other.
\end{enumerate}

Note that, in real applications, the fault features are not strictly group-sparse periodic. In other words, the real fault frequencies have a variation from the calculated frequencies (up to 1-2\%), i.e., there are not strict periods $T_1$ and $T_2$ \cite{randall2011rolling}.
In such cases, the proposed approach can still work owing to the overlapping of group structure, which will be shown in the following sections.
The proposed approach is not an improvement of the cutting edge techniques developed in the cyclostationary framework, instead, it is an alternative to these techniques.
A work closely related to the proposed approach is the periodic overlapping group sparsity (POGS) problem \cite{He_mssp_2016},
which assumes only one repetitive group-sparse component is present in the vibration signal.
Notice that although the observed signal is modeled with two components,
the proposed method also works when there exist only one fault component.
Therefore, the proposed method generalizes POGS, and is useful
when there exists multiple components.

Another related work is group-sparse signal denoising (GSSD),
which is also known as overlapping group sparsity (OGS) with non-convex regularization \cite{Chen_Selesnick_2014_GSSD},
where mathematical derivations and proofs have been given in detail to show that non-convex regularization can be used to promote group-sparsity,
while maintaining convexity of the problem as a whole.

The method proposed in this paper also uses the concept of morphological component analysis (MCA)
\cite{mca_Starck_2004},
which is a method to decompose signals based on sparse representations.
In contrast to MCA, the proposed method does not utilize any transform (e.g., Fourier or wavelet transform), i.e.,
the sparse features are in the signal domain.
Moreover, non-convex regularization is used to strongly induce sparsity while maintaining convexity of the proposed problem formulation.
The proposed approach reduces to the OGS method without prior knowledge, i.e., we can utilize the
sparsity-based OGS approach if we do not have prior knowledge of the characteristic frequencies.


\section{Preliminaries} \label{sec:preliminaries_mca}

\subsection{Notation}
In this paper, the elements of a vector $x$ are denoted as $x_n$ or $[x]_n$.
The norms of $x$ are defined as
\begin{align}
	\norm{x}_1 		: = \sum_{n} \abs{x_n}, \quad
	\norm{x}_2 	: = \bigg( \sum_{n} \abs{x_n}^2 \bigg)^{1/2}	.
\end{align}

A function of $x$ determined by parameter $a$ is denoted as $f(x ; a)$, and to distinguish it from a function with two ordered arguments, e.g., $f(x,y)$.

\subsection{Review of majorization-minimization}
In this paper, the majorization-minimization (MM) approach is used to derive a fast-converging algorithm.
This subsection briefly describes the MM approach for minimizing a convex cost function.
The MM is an approach to simplify a complicated optimization problem into a sequence of simpler ones \cite{FBDN_2007_TIP}.
More specifically, consider an optimization problem
\begin{align}
	u \opt = \arg \min_{u} F(u).
\end{align}
Using MM, the problem can be solved iteratively by
\begin{align}\label{eqn:fdmca_mm_iteration}
	u^{(i+1)} = \arg \min_{u} F\major ( u , u^{(i)} ),
\end{align}
where $F\major : \real^{N} \times \real^{N} \to \real$ is an upper bound (majorizer) of the objective function $F$, satisfying
\begin{align}\label{eqn:fdmca_mm_condition}
	&F\major(u,v) \ge F(u), \quad F\major(u,u) = F(u).
\end{align}
Note that the majorizer $F\major(u,v)$ touches $F(u)$ for $u=v$, as shown in \eqnref{eqn:fdmca_mm_condition}.
\figref{fig:fdmca_example_0_penalty} illustrates the majorizer (red line) of a penalty function (blue line), which will be described in the following subsection.
The proof of convergence for MM has been given in Ref.~\cite[Chapter~10]{mm_Lange_2000}.
More details about the MM procedure can be found in \cite{mm_Lange_2000, FBDN_2007_TIP} and references therein.

\subsection{Non-convex penalty functions}
Non-convex penalty functions can promote sparsity more strongly than convex penalty functions \cite{selesnick2014sparse,selesnick2015convex}.
This subsection briefly describes the non-convex penalty functions which will be used in the proposed approach.
The smoothed non-convex penalty function $\phie : \real \to \real_{+}$ is used in this work.
\tabref{tab:fdmca_penalty} gives several examples of the functions defined by
\begin{align}\label{eqn:fdmca_phie}
	\phie(u ; a) : = \phi( \sqrt{ u^2 +\eps } ; a ), \quad \eps > 0,
\end{align}
where $\phi$ is a non-smooth penalty function satisfying the following properties:
\begin{enumerate}
	\item
		$\phi(u;a)$ is continuous on $\real$.
	\item
		$\phi(u;a)$ is twice continuously differentiable on $\mathbb{R} \backslash \{0\}$.
	\item
		$\phi(u;a)$ is even symmetric:  $\phi(u) = \phi(-u)$.
	\item
		$\phi(u;a)$ is increasing and concave on $\real_{+}$.
	\item
		$\phi(u;a) = \abs{x}$ when $a = 0$.
\end{enumerate}
Note that, for both $\phi$ and $\phie$, the parameter $a \ge 0$ controls the concavity of the function.

\begin{figure}[t]
	\centering
    \includegraphics [scale = \Figurescale] {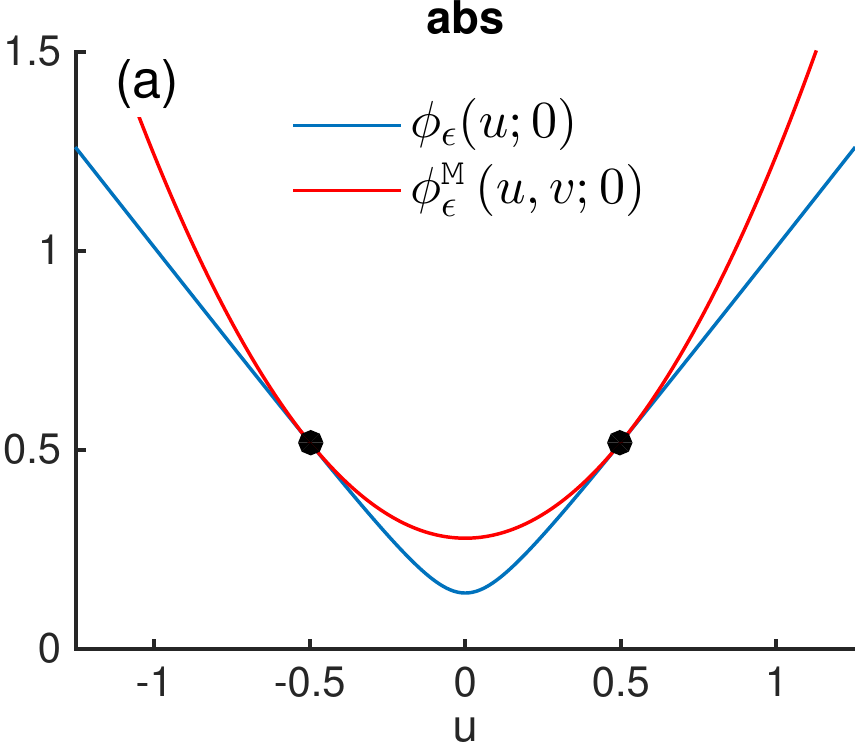}
    \quad
    \includegraphics [scale = \Figurescale] {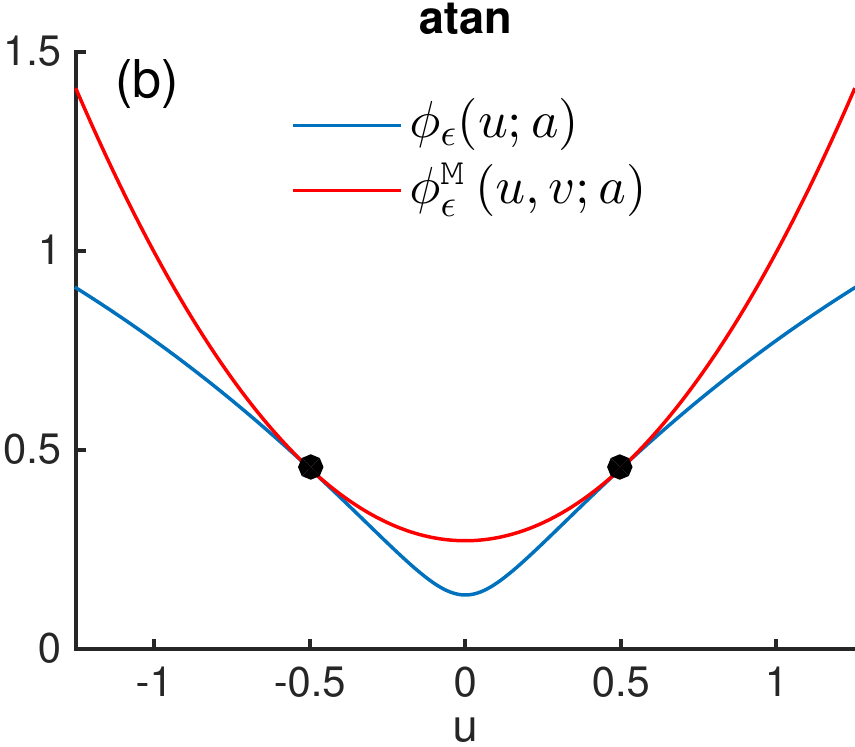}
   	\caption{
		(a) Smoothed $\ell_1$-norm (abs) penalty function (blue) and  majorizer (red).
		(b) Smoothed arctangent (atan) penalty function and majorizer (gray).
		For all the figures, the parameters are set to $v = 0.5, \eps = 0.02, a = 0.5$.		}
	\label{fig:fdmca_example_0_penalty}
\end{figure}

The parameter $\eps$ controls the smoothness of the functions.
As a special case, when $\eps = 0$, $\phie(u ; 0) = \phi(u)$, then the penalty function is non-differentiable at 0.
In practice, $\eps$ is specified very small, e.g. $10^{-10}$, so that the function is differentiable.
\figref{fig:fdmca_example_0_penalty} gives two specific examples of $\phie$.

A majorizer of penalty function $\phie$ is given by
\begin{align}\label{eqn:fdmca_phie_major}
	\phie\major( u , v ; a) := \frac{u^2}{ 2 \psi(v;a) }  -
	\underbrace{   \bigg( \frac{v^2}{2 \psi(v;a)} - \phie(v;a) \bigg) }_{\text{only depends on $v$}},
\end{align}
where $\psi(v;a)$ is listed in the third column of \tabref{tab:fdmca_penalty}. Note that $\psi(v;a)> 0$ for all $v \in \real$.
Also note that $\phie\major( u , v )$ is quadratic in $u$.
In Ref.~\cite{Ding_2015_SP}, a detailed proof has been given to show that when $\phi$ satisfies the above properties,
the majorizer \eqnref{eqn:fdmca_phie_major} satisfies the condition \eqnref{eqn:fdmca_mm_condition}
for $\phie$.

\subsection{Review of POGS}

The periodic overlapping group sparsity (POGS) problem \cite{He_mssp_2016}
considers the signal model
\begin{align}\label{eqn:fdmca_pogs_model}
	y = x + w,
\end{align}
where $x$ is a repetitive group-sparse signal, and $w$ is additive white Gaussian noise.
The POGS method extracts $x$ by solving a convex problem
\begin{align}\label{eqn:fd_pogs}
	x \opt = \arg \min_{x}
	\Big\{
		 	P_0(x) 			
    	 = \half \norm{ y - x }_2^2  + \lam \Phi (  x , b ; a )
	\Big\},
\end{align}
where $\Phi : \real^{N} \to \real$ is defined as
\begin{align}\label{eqn:fdmca_Phi}
	\Phi(x, b ; a) := \sum_n \phie \Big( \Big[\sum_k [b]_k [x]_{n+k}^2\Big]^{1/2} ; a  \Big).
\end{align}
The function $\Phi$ is a regularization term that promotes repetitive group-sparsity.
Moreover, $b \in \{ 0,1 \}^K$ is a binary weight array designated according to the period.
To simplify following derivation, denote that
\begin{align}\label{eqn:fdmca_Phi}
	\Phi(x, b; 0) = \Phi(x; b),
\end{align}
where the penalty function is strictly convex as shown in \figref{fig:fdmca_example_0_penalty}(a).

\begin{table*}[t]
\caption{Sparsity-promoting penalty functions.}
\begin{center}
	\scalebox{0.725}{
		\begin{tabular}{@{} c l l l l@{}}
		\toprule
		\textbf{Penalty}	 &$\phi(u;a)$ 	&$\phie(u;a)$  &$\psi(u;a) = u / \phie'(u;a) \quad $ \\[0.4em]
		\midrule
			abs ($a= 0$)	 	& $\abs{u}$						
	 				& $\sqrt{ u^2 + \eps }$		
	 				& $\sqrt{ u^2 + \eps }$ \\[1.2em]	
			log	 	&$\ds\frac{1}{a} \log ( 1 + a |u| )$
	 				&$\ds\frac{1}{a} \log ( 1 + a \sqrt{u^2+\eps} )$
	 				&$\ds\sqrt{ u^2 + \eps } \left( 1 + a \sqrt{ u^2 + \eps } \right)$	\\[1.0em]
			rat	 	&$\ds\frac{\abs{u}}{ 1 + a\abs{u} /2 }$
	 				&$\ds\frac{ \sqrt{ u^2 + \eps} }{ 1 + a \sqrt{ u^2 + \eps } /2 }$
	 				&$\ds\sqrt{ u^2 + \eps } \left( 1 + a \sqrt{ u^2 + \eps}/2 \right)^2$	\\[1.0em]
			atan 	&$\ds\frac{2}{a \sqrt{3}} \left( \tan^{-1} \left( \frac{1+2 a |u|} {\sqrt{3}} \right) - \frac{\pi}{6}\right)$
	 				&$\ds\frac{2}{a \sqrt{3}} \left( \tan^{-1} \left( \frac{1+2 a \sqrt{u^2+\eps} } {\sqrt{3}} \right) - \frac{\pi}{6} \right)$
	 				&$\ds\sqrt{ u^2 + \eps } \left( 1 + a \sqrt{ u^2 + \eps } + a^2 (u^2 + \eps) \right)$	\\[0.8em]
		\bottomrule
		\end{tabular}
	}
	\end{center}
	\label{tab:fdmca_penalty}	
\end{table*}

\section{Repetitive transients extraction algorithm}

In this section, the proposed algorithm termed 
\underline{r}epetitive \underline{t}ransients \underline{e}xtraction \underline{a}lgorithm (RTEA) 
is presented.

\subsection{Problem formulation}

To extract two repetitive group-sparse components, an optimization problem is  formulated as
\begin{align}\label{eqn:fdmca_problem}
	 \{ x_1\opt , x_2\opt \} & = \arg \min_{x_1 , x_2} \Big\{  P( x_1 , x_2 )
	 =   \half \norm{ y - ( x_1 + x_2 ) }_2^2 + \lam_0 R(x_1 , x_2 ; a_0)   + \sum_{i \in \{ 1,2 \}} \lam_i \Phi(x_i; b_i)  \Big\}
\end{align}
where $R : \real^{N} \times \real^{N} \to \real$ is defined as
\begin{align}\label{eqn:fdmca_R}
	R( x_1 , x_2 ; a_0) 	& := \sum_{n}  \phie \Big( \Big[ \sum_{k=0}^{K_0-1} [x_1+x_2]_{n+k}^2 \Big]^{1/2}; a_0 \Big) .
\end{align}
The function $R$ is an overlapping group sparsity (OGS) regularization function with group size $K_0$%
\footnote{Using the notation of \cite{He_mssp_2016},
it is denoted $ \sum_i \phie( \norm{x}_{i,K_0} ; a_0 ) $. }.
There are two more regularizers in \eqnref{eqn:fdmca_problem}
promoting the repetitive group-sparsity of $x_1$ and $x_2$ respectively, and formulated in \eqnref{eqn:fdmca_Phi}.
%

Furthermore, in problem \eqnref{eqn:fdmca_problem}, $b_1 \in \{ 0,1 \}^{K_1}$  and $b_2 \in \{ 0,1 \}^{K_2}$
are two binary-weighting arrays as
\begin{align} \label{eqn:fdmca_preoid_b12}
	b_i = [ \ \underbrace{
				  \underbrace{ 1 ~ 1 \cdots 1 }_{N_{i1}}
				\ \underbrace{ 0~0 \cdots 0 }_{N_{i0}}
				\	\dots
				\ \underbrace{ 1~1 \cdots 1 }_{N_{i1}}
				\ \underbrace{ 0~0 \cdots 0 }_{N_{i0}}
				\ \underbrace{ 1~1 \cdots 1 }_{N_{i1}}}_{ \text{spanning } M_i \text{ periods}}  \
			],
\end{align}
for $i = 1,2$ and $M_1$ and $M_2$ defines the number of periods included in $b_1$ and $b_2$ respectively.

Moreover, in contrast to MCA, which has two regularizers,
problem \eqnref{eqn:fdmca_problem} has three.
The regularization term $R$ is introduced because according to the signal model,
the summation of the two components is also sparse.
Note that, $( x_1 + x_2 )$ might be sparse when $x_1$ and $x_2$ are not,
but in \eqnref{eqn:fdmca_problem}, the regularizers with $\Phi$ does force $x_1$ and $x_2$  to be sparse.

\subsection{Convexity of the objective function}

The idea of using non-covex regularization with ``maximizing concavity'' in a convex problem
has been illustrated in \cite{selesnick2014sparse},
wherein the quadratic formulation of data fidelity term can be used to compensate the non-convexity
in the regularization so that the objective function is still convex.
In this work, we adopt this idea for an extraction algorithm with simultaneous denoising.
Entirely there are three regularizers in \eqnref{eqn:fdmca_problem},
and we allow one of them to be non-convex to promote the global sparsity more strongly.
Moreover, the specific condition to assure the convexity of problem \eqnref{eqn:fdmca_problem} is derived as the following proposition.

\begin{proposition} \label{prop:fdmca_convex}
Suppose the parameterized penalty function $ \phie $ is defined by formula \eqnref{eqn:fdmca_phie} and $ \lam_0 > 0 $.
If
\begin{equation}\label{eqn:fdmca_convex}
	0 \le a_0 < \frac{1}{K_0\lam_0},
\end{equation}
then the objective function $ P : \real^{N} \times \real^{N} \to \real $ in \eqnref{eqn:fdmca_problem} is strictly convex.
\end{proposition}

A proof of the above proposition is given in \appref{app:fdmca_convex}.

\section{Algorithm derivation}

In this section, an algorithm is derived to solve \eqnref{eqn:fdmca_problem} based on MM.
The majorizer of $R$ in \eqnref{eqn:fdmca_R} is
$R\major : \real^{N} \times \real^{N} \times \real^{N} \times \real^{N} \to \real$, and written explicitly as
\begin{align}\label{eqn:fdmca_R_major}
	R& \major ( x_1, x_2, z_1, z_2 ; a_0 ) \nonumber \\					
		 	 & = \sum_n \Big\{ r_0(n, z_1+z_2)([x_1]_n^2 + [x_2]_n^2
		 - [z_1-z_2]_n [x_1]_n - [z_2-z_1]_n [x_2]_n ) \Big\} + C(z_1,z_2),
\end{align}
where $C(z_1,z_2)$ is a constant only dependent on $z_1$ and $z_2$.
In \eqnref{eqn:fdmca_R_major}, $r_0$ is a function $r_0 : \mathbb{Z} \times \real^{N} \to \real$ given by
\begin{align}
	r_0( n, z_1+z_2 ) = \sum_{j=0}^{K_0-1}{\psi^{-1} \Big( \Big[  \sum_{k=0}^{K_0-1} [z_1+z_2]_{n-j+k}^2\Big]^{1/2} ; a_0 \Big) }.
\end{align}

The majorizer of function $\Phi$ has been derived in Ref.~\cite[Section~3.3]{He_mssp_2016}.
Here,  it can be rewritten using a slightly different notation, that $\Phi \major : \real^{N} \times \real^{N} \to \real$ is
\begin{align}
	\Phi \major( x, z; b, a ) = \half \sum_n r(n,z) x_n^2 + C(z),
\end{align}
where $r: \mathbb{Z} \times \real^{N} \to \real$ is defined as
\begin{align}\label{eqn:fdmca_r}
	r(n,z) : = \sum_{j=0}^{K-1} \frac{ [b]_j }{ \psi\Big( \Big[\sum_k [b]_k [z]_{n-j+k}^2\Big]^{1/2} ; a  \Big) }
\end{align}

Using the above results, the majorizer of the objective function $P$ in \eqnref{eqn:fdmca_problem} can be written as
\begin{align}\label{eqn:fdmca_PM}
	P \major(x_1,x_2,z_1,z_2)
	= 	& \half \norm{ y - (x_1 +x_2) }_2^2 + \half \norm{ (x_1 -z_1) - (x_2 -z_2) }_2^2 \nonumber \\
		& \quad + \lam_0 R\major(x_1,x_2,z_1,z_2;a_0) + \! \sum_{i\in\{1,2\}} \! \lam_i \Phi \major( x_i, z_i; b_i, a_i ),
\end{align}
where the function $R\major : \real^{N} \times \real^{N} \times \real^{N} \times \real^{N} \to \real$
is a majorizer of $R$ in \eqnref{eqn:fdmca_R},
wherein the derivation in detail is given in \appref{app:fdmca_R_func}.

\begin{table}[t]
\caption{Explicit steps of proposed method (RTEA).}
	\begin{empheq}[box=\fbox]{align*}
		& \text{Input:} ~y \in \real^N,~K_0, ~\lam_0, ~\lam_i, ~b_i\in \{ 0, 1 \}^{K_i}, \text{ for } i \in \{ 1,2 \}.		\nonumber		\\
		& \text{Initialization: } a_0\ge0, ~a_i\ge0, ~x_i = y, \text{ for } i = 1,2.  \\
		& \text{Repeat:}  \\
		& \quad [r_0]_n = \sum_{j=0}^{K_0-1} \frac{1} {\displaystyle \psi_0 \Big( \Big[  \sum_{k=0}^{K_0-1} [x_1+x_2]_{n-j+k}^2\Big]^{1/2} ; a_0 \Big) } \\
		& \quad [r_i]_n = \sum_{j=0}^{K_i-1} \frac{ [b_i]_j }
			{ \displaystyle\psi_i\Big( \Big[ \sum_{k=0}^{K_i-1}  [b_i]_k [x_i]_{n-j+k}^2\Big]^{1/2} ; a_i  \Big) }, \text{ for } i \in \{ 1,2 \} \\[0.4em]
		& \quad [p_i]_n = 2 + 2\lam_0 [r_0]_n   + \lam_i [r_i]_n , \text{ for } i \in \{ 1,2 \}\\
		& \quad [q_1]_n = y_n + ( 1+[r_0]_n )[x_2-x_1]_n	\\
		& \quad [q_2]_n = y_n + ( 1+[r_0]_n )[x_1-x_2]_n	\\[0.4em]
		& \quad [x_i]_n = [q_i]_n/[p_i]_n, \text{ for } i \in \{ 1,2 \} \\
		& \text{Until convergence} \\
		& \text{Return: } x_1, x_2
	\end{empheq}\label{alg:fdmca_main}
\end{table}

Note that in \eqnref{eqn:fdmca_PM}, an extra term
$ \half \norm{ (x_1 -z_1) - (x_2 -z_2) }_2^2 $ is introduced, which does not break the property of majorizer,
but helps to cancel the term $x_1^{\tp}x_2$ in the data fidelity term.
Then $P \major(x_1,x_2,z_1,z_2) $ can be written as
\begin{align}\label{eqn:fdmca_PM_2}
	P \major(x_1,x_2,z_1,z_2)  =
	& \sum_n  \half [ p_1(z_1,z_2) \odot x_1]_n^2 + \half [ p_2(z_1,z_2) \odot x_2]_n^2	\nonumber \\
	& - [ q_1(z_1,z_2) \odot x_1]_n - [ q_2(z_1,z_2) \odot x_2]_n + C(z_1,z_2),
\end{align}
where $\odot$ denotes element-wise multiplication,
and the coefficients $p_1,p_2, q_1,q_2 \in \real^{N}$ can be written explicitly as
\begin{subequations}
\begin{align}
	[p_1(z_1,z_2)]_n &= 2 + 2\lam_0 r_0(n,z_1+z_2)   + \lam_1 r_1(n,z_1) \\
	[p_2(z_1,z_2)]_n &= 2 + 2\lam_0 r_0(n,z_1+z_2)   + \lam_2 r_2(n,z_2) \\
	[q_1(z_1,z_2)]_n &= y_n + [ 1+r_0(n,z_1+z_2) ][z_1-z_2]_n			\\
	[q_2(z_1,z_2)]_n &= y_n + [ 1+r_0(n,z_1+z_2) ][z_2-z_1]_n.
\end{align}
\end{subequations}
Note that $r_1$ and $r_2$ are functions defined by \eqnref{eqn:fdmca_r} dependent on
the binary weighting arrays $b_1$ and $b_2$  in \eqnref{eqn:fdmca_preoid_b12} respectively.

\begin{figure}[t!]
	\centering
    \includegraphics [scale = 0.8] {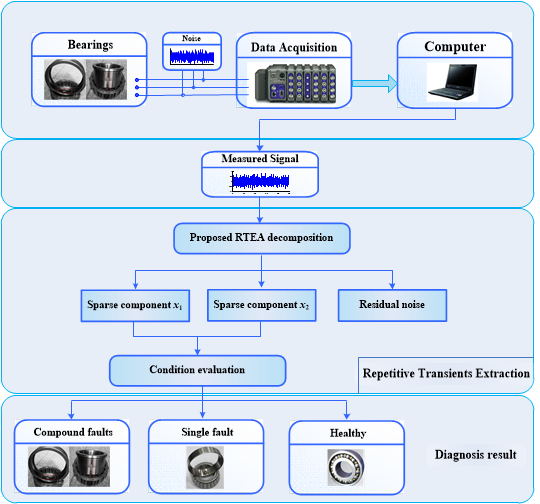}
   	\caption{Procedure of the proposed RTEA for fault detection of bearings.}
	\label{fig:mca_procedure}
\end{figure}

Consequently, using MM, the problem \eqnref{eqn:fdmca_problem} is solved iteratively by
\begin{align}\label{eqn:fdmca_mm_iteration}
	\{ x_1^{(i+1)} , x_2^{(i+1)} \}  = \arg \min_{x_1,x_2} P\major ( x_1,x_2,  x_1^{(i)} , x_2^{(i)} ),
\end{align}
and the results of $x_1$ and $x_2$ in each iteration can be written separately as
\begin{subequations}
\begin{align}
	[x_1^{(i+1)}]_n & = \frac{ [q_1( x_1^{(i)} , x_2^{(i)})]_n } {[ p_1( x_1^{(i)} , x_2^{(i)})]_n }, \\
	[x_2^{(i+1)}]_n & = \frac{ [q_2( x_1^{(i)} , x_2^{(i)})]_n } {[ p_2( x_1^{(i)} , x_2^{(i)})]_n }.
\end{align}
\end{subequations}
\tabref{alg:fdmca_main} gives the specific steps of the proposed algorithm RTEA.
Note that, in this algorithm, the computation of $r_0$, $r_1$ and $r_2$ can be implemented directly with convolution,
and the rest steps are element-wised independent,
where practically multi-threading computation can be directly adopted.
The systematic structure of the proposed RTEA for fault detection of rolling element bearings is presented in \figref{fig:mca_procedure}.
Since this algorithm is derived using MM procedure, the convergence can be guaranteed and converges to the optimal minimizer in this case.
The detailed proof of MM for convex problems can be found in Ref.~\cite[Chapter~10]{mm_Lange_2000}.

\begin{figure}[t!]
	\centering
    \includegraphics [scale = \figurescale] {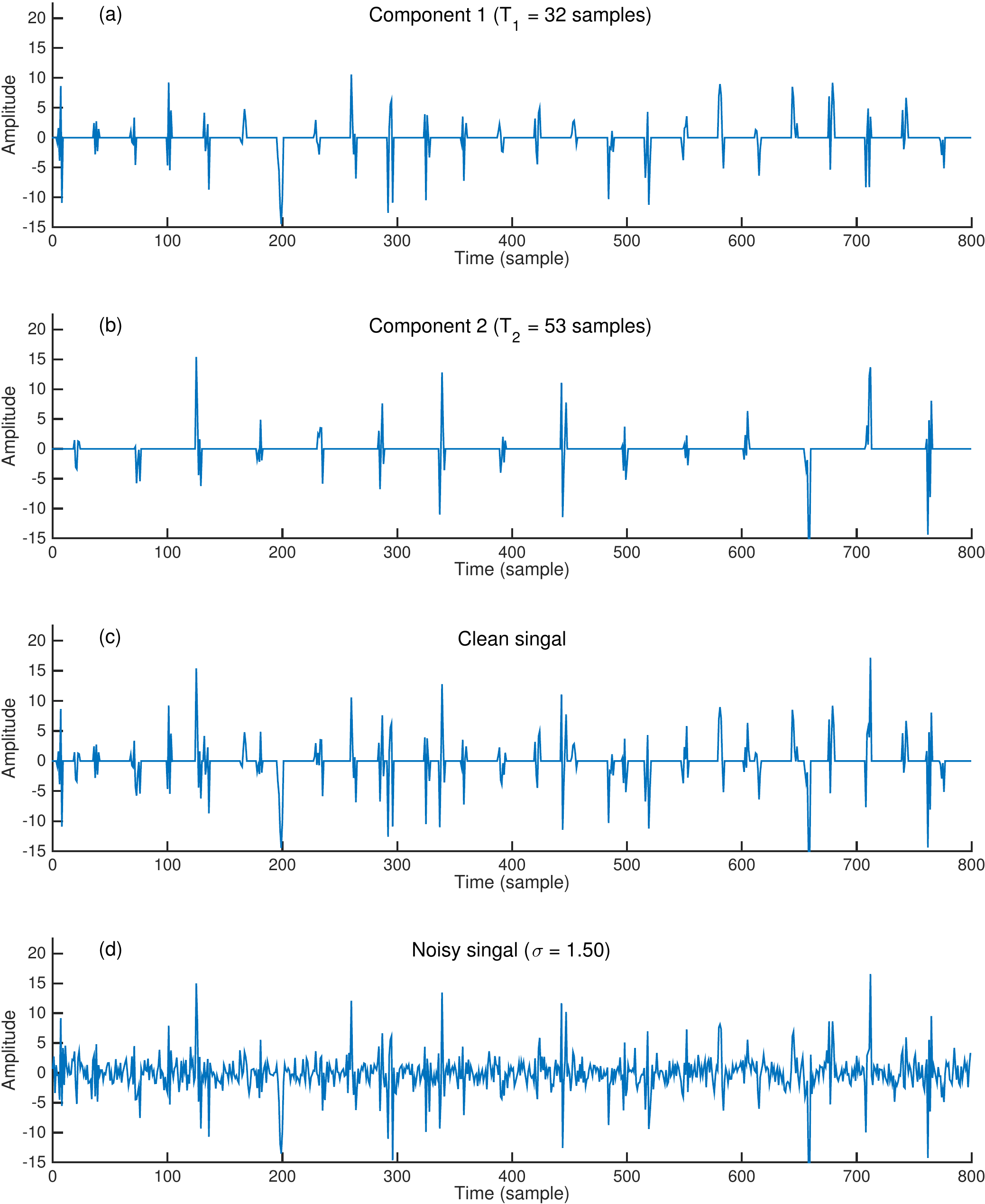}
    \caption{Example~1: Test signal.
		(a) Component 1 with period $T_1 = 32$ samples.
		(b) Component 2 with period $T_2 = 53$ samples.	
		(c) Summation of the two compoennts.
		(d) Noisy observation.
		}
	\label{fig:fdmca_example_1_test}
\end{figure}

\begin{figure}[t!]
	\centering
    \includegraphics [scale = \figurescale] {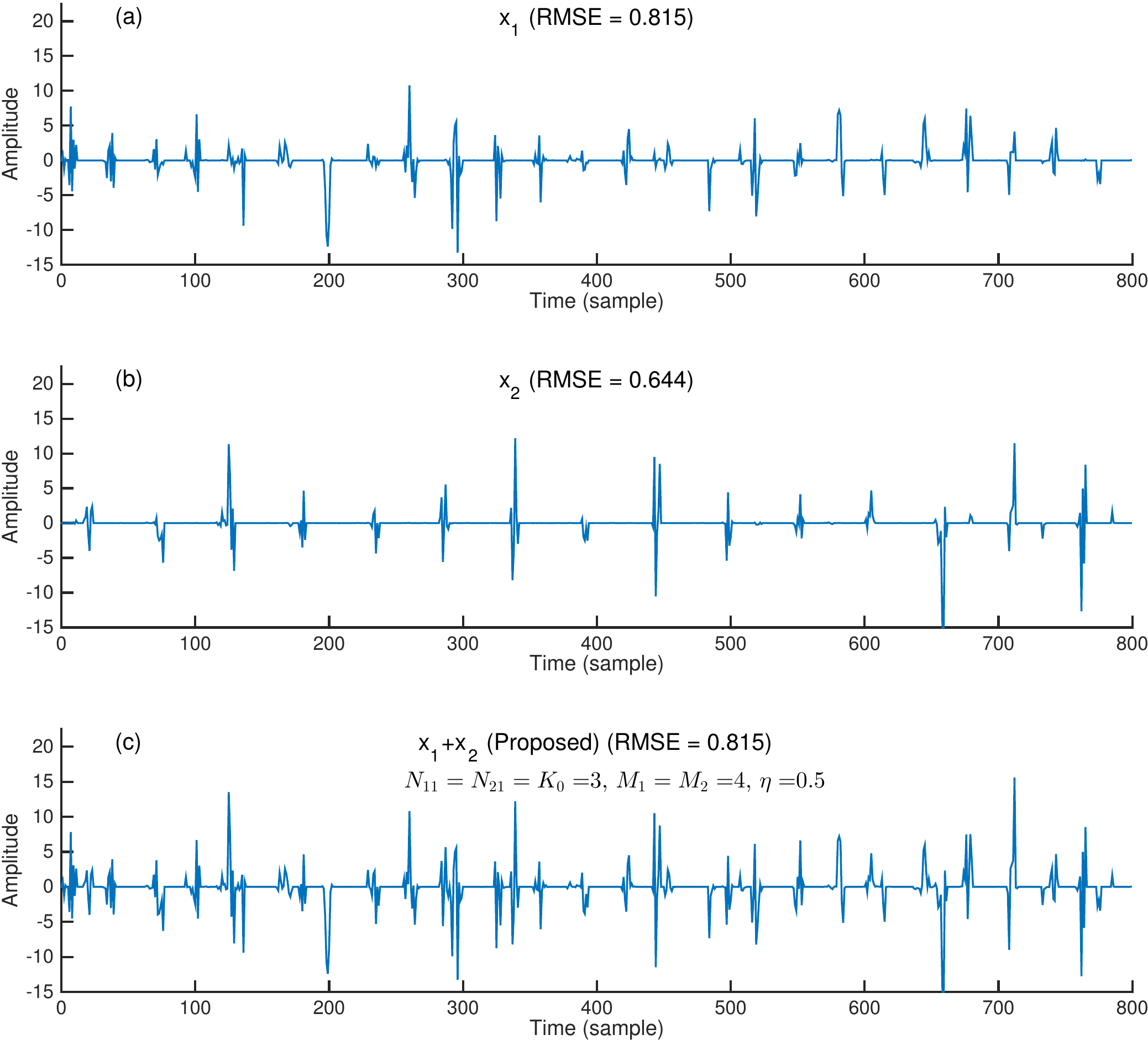}
       	\caption{Example~1: Results of proposed method.}
	\label{fig:fdmca_example_1_result}
\end{figure}

\begin{figure}[t!]
	\centering
    \includegraphics [scale = \figurescale] {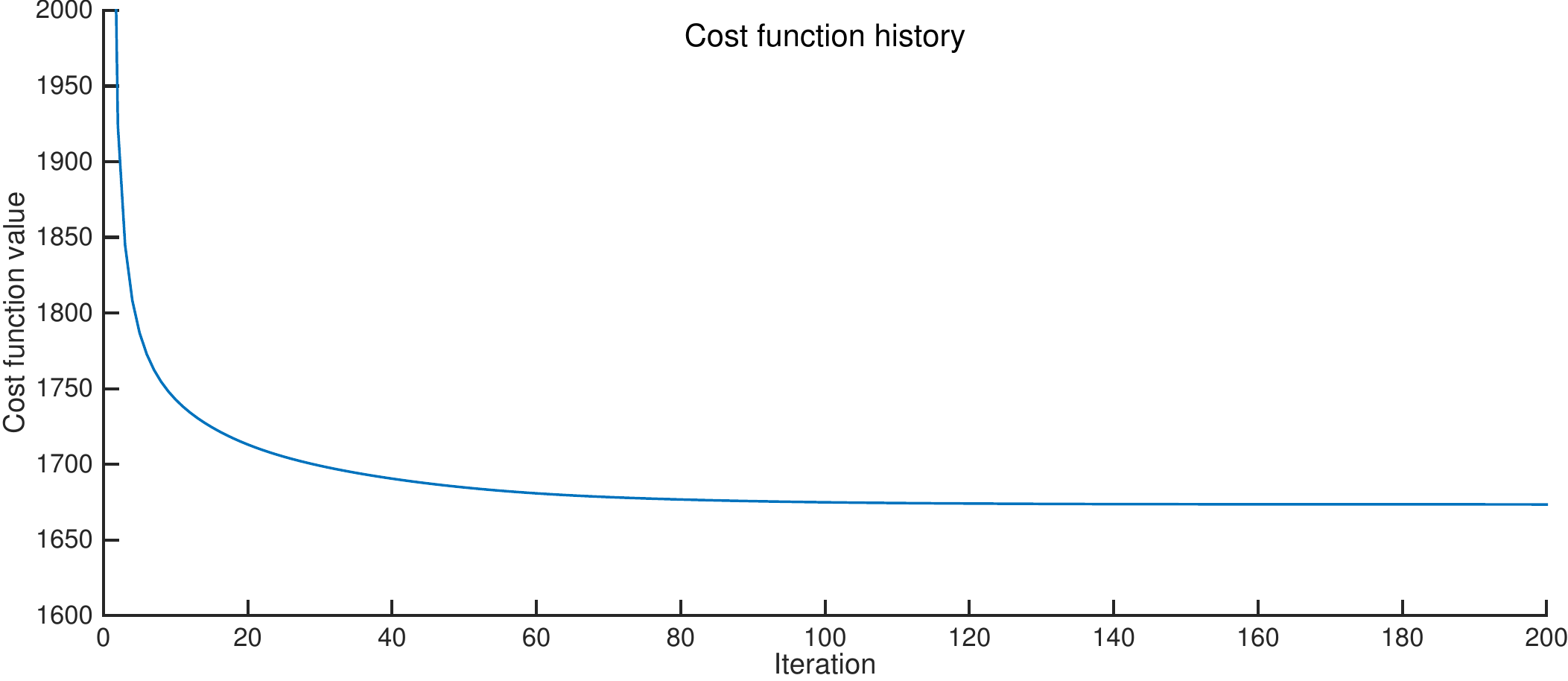}
       	\caption{Example~1: Cost function history.}
	\label{fig:fdmca_example_1_cost}
\end{figure}

\section{Synthetic data example}

\figref{fig:fdmca_example_1_test} shows the synthetic test signal.
It consists of two sparse transient sequences exhibiting distinct periods, where $T_1 = 32$ and $T_2 = 53$ samples.
In this example, each transient is generated by adding a random number of sinusoids with random frequencies and initial phases.
More specifically, each transient can be written as
\begin{align}\label{eqn:fdmca_transient}
	g(n) = \sum_{j=1}^{J} A_j \sin( \omega_j n + \theta_j ), \quad n \in \{ 0,1,2 \dots 9 \},
\end{align}
where $1 \le J \le 10$ is a random integer,
and for each $j$, $A_j$ is a random amplitude,
$\omega_j $ is a random frequency,
and $\theta_j $ is a random phase.
The sequences of transients are shown in \figref{fig:fdmca_example_1_test}(a) and \figref{fig:fdmca_example_1_test}(b),
and their summation is shown in \figref{fig:fdmca_example_1_test}(c).

In this example, the parameters are set to $N_{11} = N_{21} = K_0 = 3$ and $M=4$,
to determine the binary arrays $b_1$ and $b_2$ by the priori known periods.
The root-mean-square error (RMSE) is used as an evaluation metric.
Moreover, the penalty functions are selected by Proposition~\ref{prop:fdmca_convex} assuring the problem is strictly convex.
\figref{fig:fdmca_example_1_result} shows the results from the proposed method,
and \figref{fig:fdmca_example_1_cost} shows the convergence behavior of the proposed algorithm.

\begin{figure}[t!]
	\centering
    \includegraphics [scale = \figurescale] {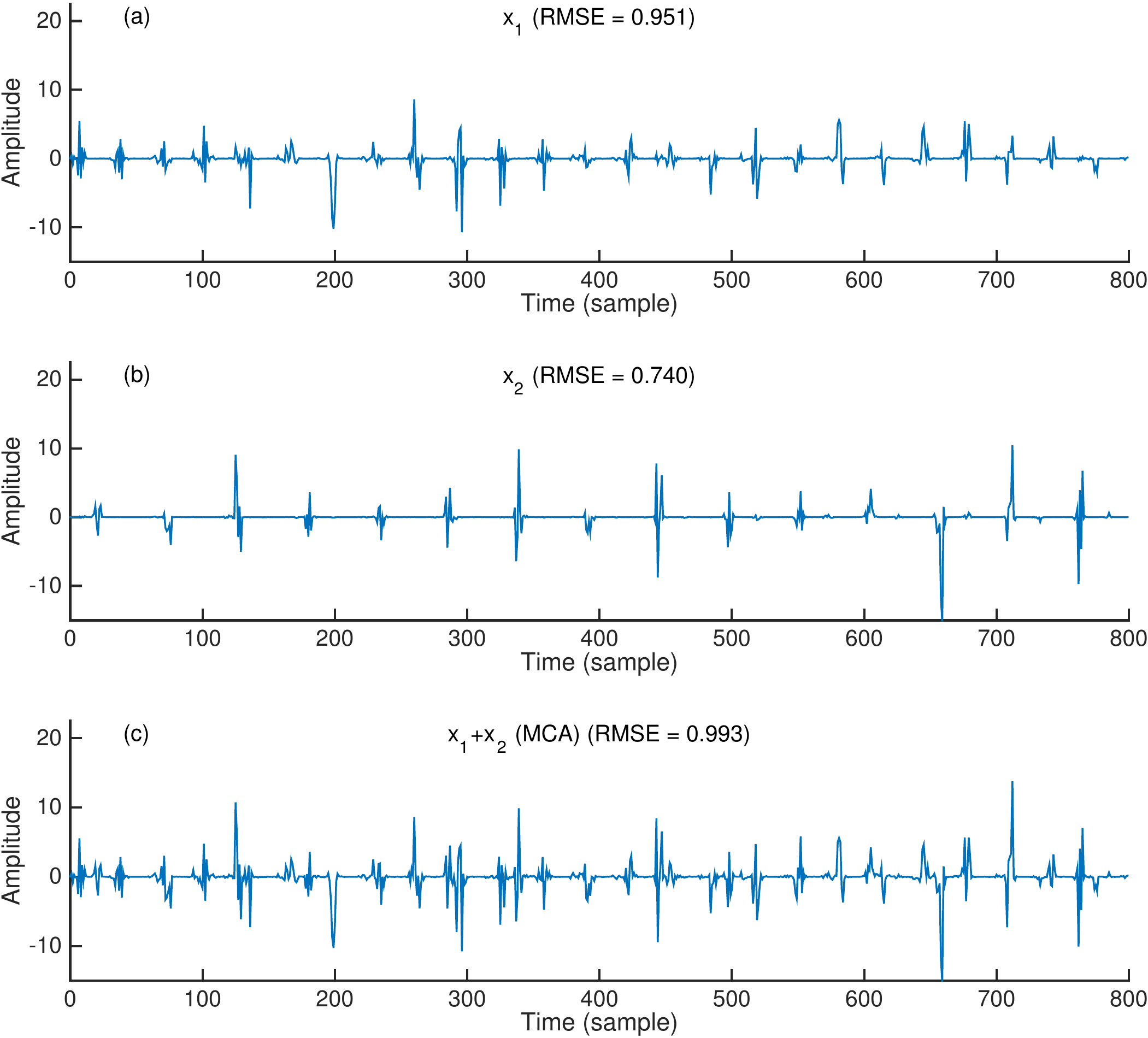}
       	\caption{Example~1: Results by using \eqnref{eqn:fdmca_mca} with convex regularizations.}
	\label{fig:fdmca_example_1_mca_result}
\end{figure}

\begin{figure}[t!]
	\centering
    \includegraphics [scale = \figurescale] {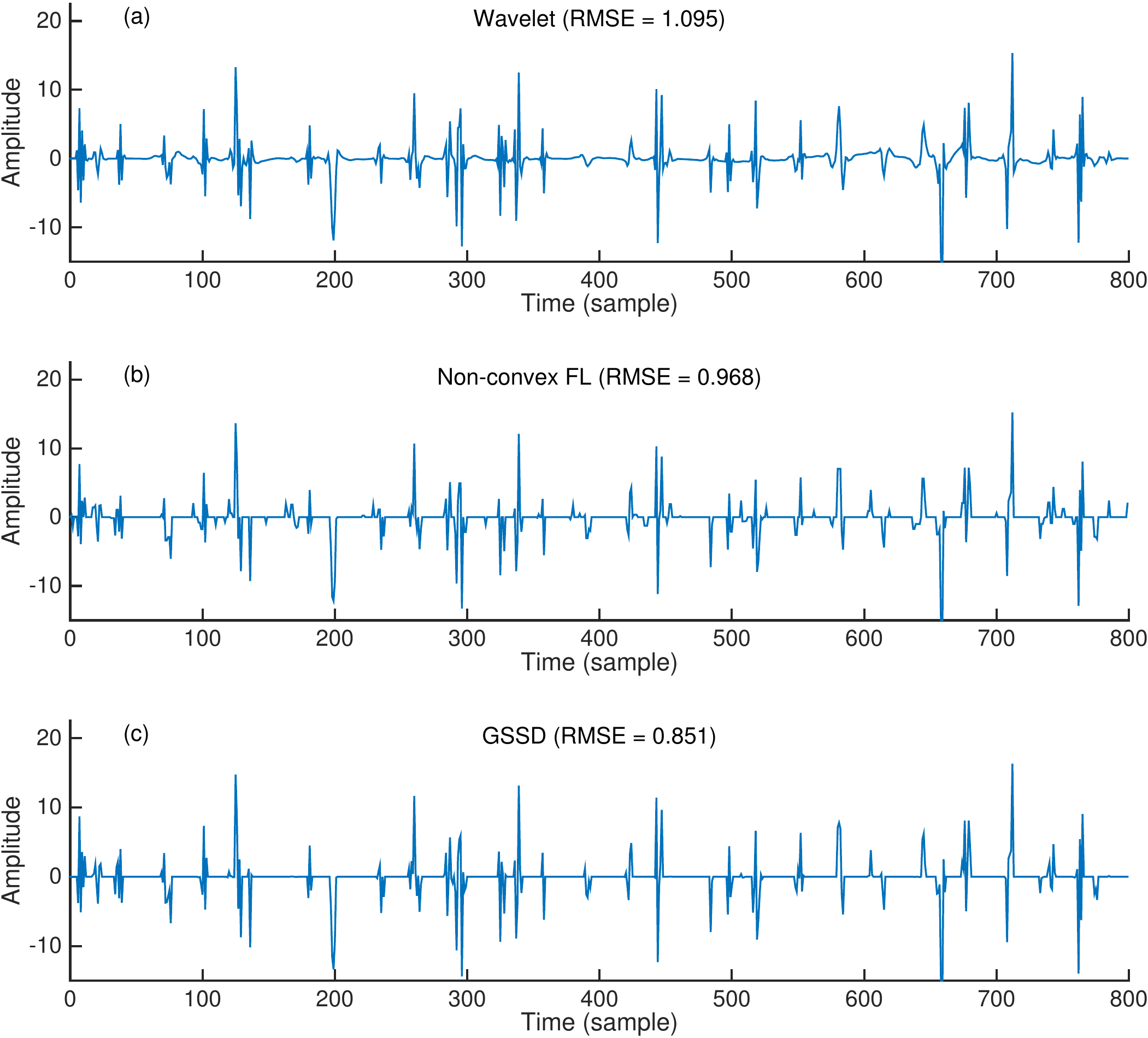}
    \caption{Example~1: Results by using some popular and/or state-of-the-art methods.
		(a) Undecimated wavelet	based denoising.
		(b) Fused lasso with non-convex regularization.
		(c) group-sparse signal denoising (GSSD) with group size 3.}
	\label{fig:fdmca_example_1_other_result}
\end{figure}

As a comparison, considering another problem,
based strictly on MCA, whose objective function is
\begin{align}\label{eqn:fdmca_mca}
\arg \min_{x_1 , x_2} \half \norm{ y - ( x_1 + x_2 ) }_2^2
	+ \sum_{i \in \{1,2\}}\lam_i \Phi(x_i; b_i).
\end{align}
Note that \eqnref{eqn:fdmca_mca} has one regularization term less than \eqnref{eqn:fdmca_problem},
where the global sparsity is not promoted by function $R$.

Although non-convex penalties can help to promote sparsity,
this will break the convexity of problem \eqnref{eqn:fdmca_mca}, so that a global optimal solution is not assured.
Moreover, experiment results find that convex formulation \eqnref{eqn:fdmca_problem}
obtains as good a performance as the non-convex formulation \eqnref{eqn:fdmca_mca}.
As a consequence, the formulation \eqnref{eqn:fdmca_problem} is preferred, because there is no need to sacrifice convexity.

In addition, the proposed method with non-convex formulation is also evaluated,
where $a_1$ and/or $a_2$ are greater than zero.
In this case, losing the convexity, the resulting sparsity can be even further promoted,
then the formulation \eqnref{eqn:fdmca_problem} is still preferred to the MCA formulation \eqnref{eqn:fdmca_mca}.

Further comparisons to some denoising methods are also presented.
Firstly, wavelet-based denoising method is adopted to the test signal,
more specifically, a 6-scale undecimated wavelet transform \cite{Coifman_1995} using Haar wavelet filter.
For denoising, hard-thresholding is applied and the threshold value is chosen by $3\sigma$-rule for each subband.
As shown in \figref{fig:fdmca_example_1_other_result}(a), the denoising results adheres the shape of impulse response of the wavelet,
where some of the signal does not have a zero baseline.

Secondly, the non-convex regularized fused lasso (FL) approach proposed in \cite{Bayram_ncvx_icassp_2014} is adopted,
which is an improved version of conventional FL \cite{Tibshirani_2005},
allowing one regularizer to be non-convex and preserving the global convexity.
The result is shown in \figref{fig:fdmca_example_1_other_result}(b).
Thirdly, group-sparse signal denoising (GSSD) also known as OGS with non-convex regularization \cite{Chen_Selesnick_2014_GSSD}
is adopted.
More specifically, `atan' penalty with group size to be 3 samples is used,
and the regularization parameter is chosen to optimize the RMSE.
The result is shown in \figref{fig:fdmca_example_1_other_result}(c).
Both of the above methods also have worse recoveries in terms of RMSE.
In addition, these methods are all denoising methods only,
they cannot decompose the signal into two distinct sequences of transients, each exhibiting its own period.

\subsection{Parameter selection}

\begin{table}[h!]
	\caption{Selection of $\beta_i$ for $i = 0,1,2$.} 
	\label{tab:fdmca_beta}
\begin{center}
    {
		\begin{tabular}{ l  c cccc }
		\toprule
		\backslashbox{$M_i$}{$N_{i1}$}	 &1   &2 &3 &4 \\[0.2em]
		\midrule
		~1 $(K_0)$	&3.700   &1.700 &1.150 &0.925	\\
		~2 	 		&1.700   &0.850 &0.625 &0.475	\\
		~3 	 		&1.150   &0.625 &0.450 &0.375	\\
		~4 	 		&0.925   &0.475 &0.375 &0.325	\\[0.1em]
		\bottomrule
		\end{tabular}
	}
	\end{center}
\end{table}

In Section~4.1 of \cite{He_mssp_2016},
the schemes to set the binary weight array $b$ \eqnref{eqn:fdmca_preoid_b12}
and the regularization parameter $\lam$ for POGS problem have been discussed in detail.
Moreover, a look-up table has been given as a guide to choose the regularization parameter,
where using the given multiplier in \cite[Table~3]{He_mssp_2016}, the regularization parameter
can be chosen by
	$\lam = \beta \sigma_w$,
where $\sigma_w$ is the deviation of the additive noise.
Here, the table is quoted with a slight change of notation.

\begin{figure}[t!]
	\centering
    \includegraphics [scale = \figurescale] {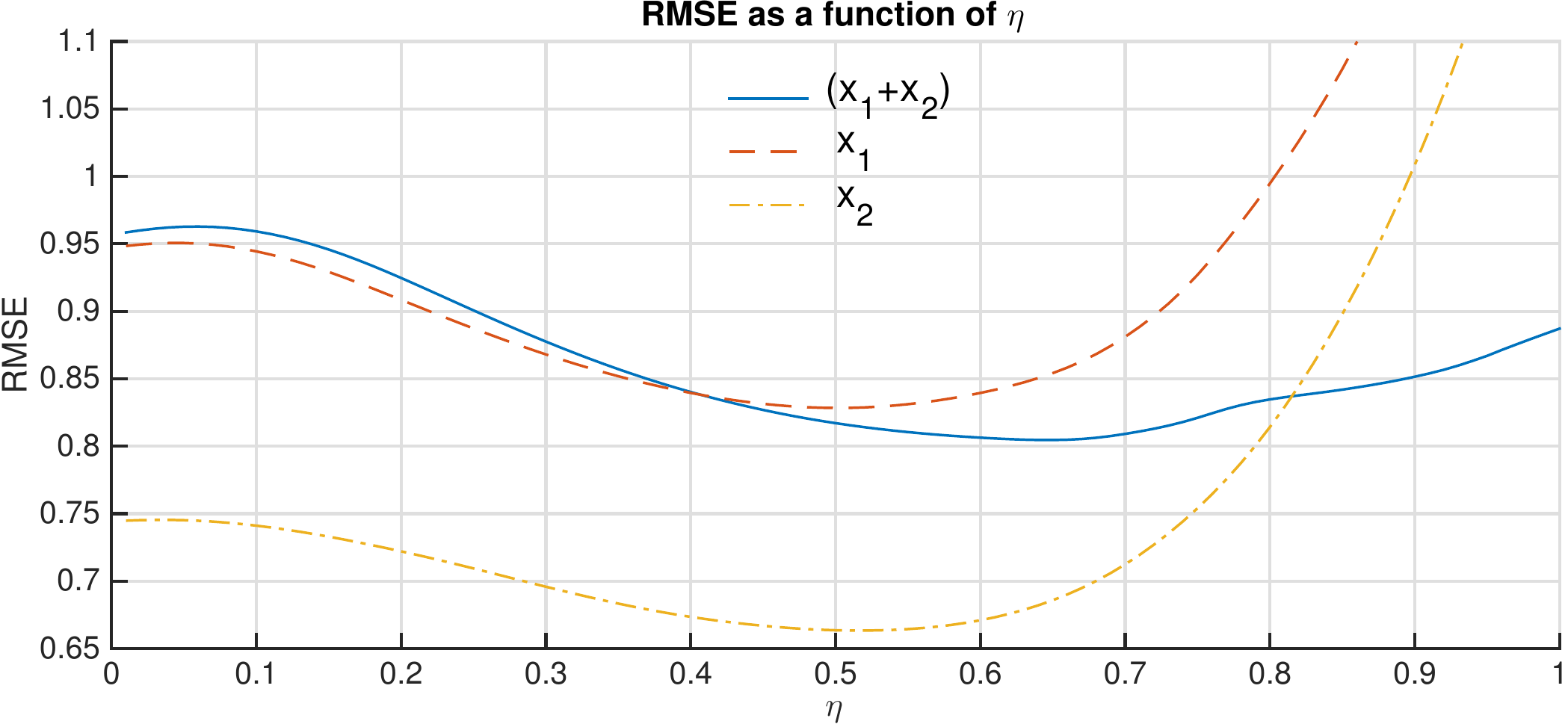}
    \caption{Example~1: RMSE values as a function of $\eta$ using \eqnref{eqn:fdmca_lam}.}
	\label{fig:fdmca_example_1_eta}
\end{figure}

As an extension of POGS with simultaneously decomposition,
this parameter selection scheme can be used
with a modification wherein the weights are shared among the three regularizers.
In this case, the regularization parameters $\{ \lam_0,  \lam_1, \lam_2 \}$ are suggested to be set by
\begin{subequations}\label{eqn:fdmca_lam}
\begin{align}
	\lam_0 & = \eta \beta_0 		\label{eqn:fdmca_lam_a}\\
	\lam_i & = 0.5(1-\eta) \beta_i  ~\text{ for } i = 1,2 \label{eqn:fdmca_lam_b}
\end{align}		
\end{subequations}
where $ 0 < \eta < 1 $ is a parameter to balance sparsity of the sum $(x_1 + x_2)$ and sparsity of the individual signal component $x_i$.

To run the method,
firstly the binary weight arrays $b_1,b_2$ need to be set by \eqnref{eqn:fdmca_preoid_b12}
using the priori periods.
Then $\beta_i$ can be determined by \tabref{tab:fdmca_beta} using $N_{i1}$ and $M_i$
for $i = 1,2$.
Note that it is necessary to chose $K_0 = \min\{ N_{11}, N_{21} \} $ to induce group-sparsity.
Hence in practice $K_0$ is not necessary to be chosen, and so as $\beta_0$.

In practice, the parameters are set to be $K_0 = N_{11} = N_{21}$, and $M_1 = M_2 = 4$.
The regularization parameters can be straight-forwardly determined by \tabref{tab:fdmca_beta} and \eqnref{eqn:fdmca_lam}.

\medskip
\textbf{Setting parameter $\eta$}.
The parameter $ 0 < \eta < 1 $ in \eqnref{eqn:fdmca_lam} is to balance the sparsity, as mentioned above.
As special cases:
\begin{enumerate}
\item
	If $\eta \to 1$, then $ \lam_1=\lam_2 \to 0 $,
	and the approach promotes sparsity of $x_1 + x_2$, but leads to $x_1 = x_2$.
\item
	If $\eta \to 0$, then the problem \eqnref{eqn:fdmca_problem} reduces to \eqnref{eqn:fdmca_mca},
	which is a conventional MCA problem. It is able to seperate the components,
	but promotes sparsity weakly.
\end{enumerate}
To obtain both of the benefits, through experiments, parameter $\eta$ is suggested to be set around 0.5.

\figref{fig:fdmca_example_1_eta} illustrates the RMSE values of $x_1,x_2$ and $(x_1+x_2)$ as functions of $\eta$.
As shown, when $\eta$ is greater than 0.5, although the RMSE of the global signal $(x_1+x_2)$ is low,
the performance of the decomposition is worse.
Moreover, $\eta$ should not be too small,
for then the results will be similar to conventional MCA shown in \figref{fig:fdmca_example_1_mca_result}.

\section{Engineering Examples}

\begin{figure}[h]
	\centering
    \includegraphics [scale = 0.8] {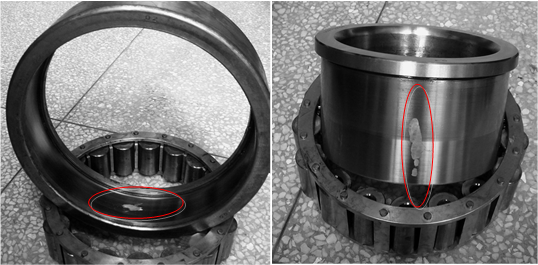}
   	\caption{Outer race defect (left) and inner race (right) in the bearing.}
	\label{fig:bearing_fault}
\end{figure}

\begin{table}[h]
  \centering
    \caption{Parameters of 552732QT bearing} 
  \begin{tabular}{@{} ccccc @{}} 
    \toprule
    Inner Race (mm) & Outer Race(mm) & Roller (mm) & Number of rollers & Contact angle (degree) \\
    \midrule
    160 & 290 & 34 & 17 & $0$ \\
    \bottomrule
  \end{tabular}
  \label{tab:example_2_parameters}
\end{table}

In this section, the proposed RTEA is applied to analyze vibration signals collected from a rolling element bearing with compound defects on a locomotive,
using a SONY~EX data acquisition system operating at a sampling rate of $f_s = 12.8$ kHz.
The locomotive bearing with faults in inner and out races is shown in \figref{fig:bearing_fault}.
The bearing (552732QT) parameters are given in \tabref{tab:example_2_parameters}.

\subsection{Compound faults detection}

In this example,
the vibration signals were collected at a constant shaft speed of 360 r/min.
Thus, based on the geometric parameters and rotational frequency,
the ball-pass frequency of the outer race is about $43.3$ Hz,
and that of the inner race is about $58.7$ Hz.

In practice, the regularization parameters in \eqnref{eqn:fdmca_problem} can be estimated from the noise level,
or the deviation of the vibration signal without any fault under an approximately identical experiment environment.
Moreover, when the healthy data is not available, the `noise' level can still be determined by the formula
\begin{align}\label{eqn:fdmca_mad}
	\hat \sigma = \mathsf{MAD}(y) / 0.6745
\end{align}
which is a conventional estimator of noise level used for wavelet-based denoising \cite{wav_Donoho_93_ideal},
where $ \mathsf{MAD}$ is the median absolute deviation defined as
\begin{align}\label{eqn:fdmca_mad_2}
	\mathsf{MAD}(y)  :=  \mathsf{median} ( \abs{  y_n - \mathsf{median}(y)  }).
\end{align}
In this example, the formula \eqnref{eqn:fdmca_mad} is used to estimate $\sigma$ directly from the observation data,
and then that value is used to choose the regularization parameters from \tabref{tab:fdmca_beta}.

\begin{figure}[t]
	\centering
    \includegraphics [scale = \figurescale] {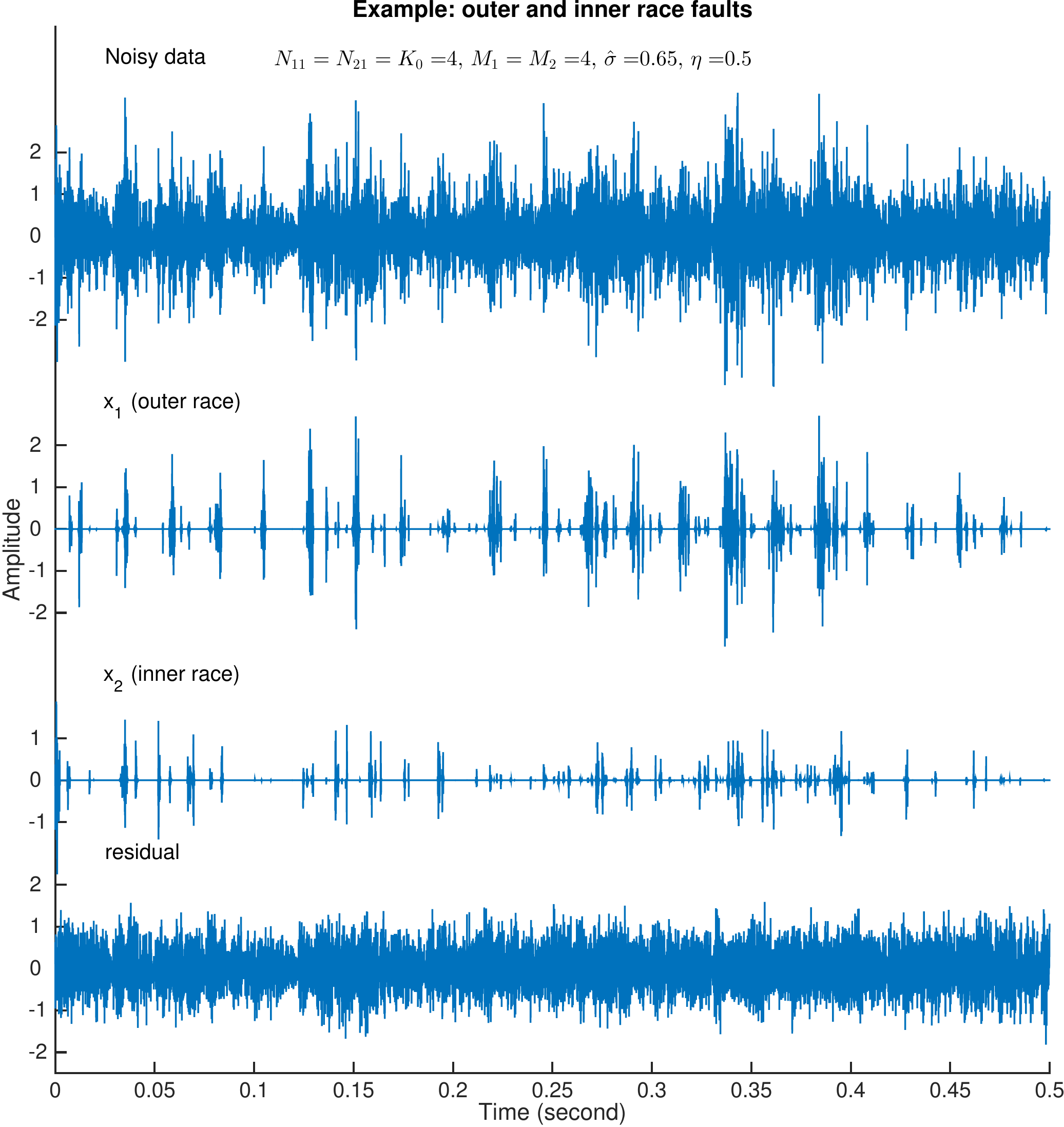}
    \caption{Example~2: Output of RTEA (proposed method) from data with compound faults.}
	\label{fig:fdmca_example_2_result}
\end{figure}

\figref{fig:fdmca_example_2_result} shows the results using RTEA,
where $x_1$ is the transient sequence generated by the fault in the outer race,
and $x_2$ is the transient sequence generated by the fault in the inner race.
During the recording, the outer race is stable and the inner race is rotating.
Hence, the amplitude of transients in $x_2$ (inner race fault) exhibit modulating effect.

\begin{figure}[t!]
	\centering
    \includegraphics [scale = \figurescale] {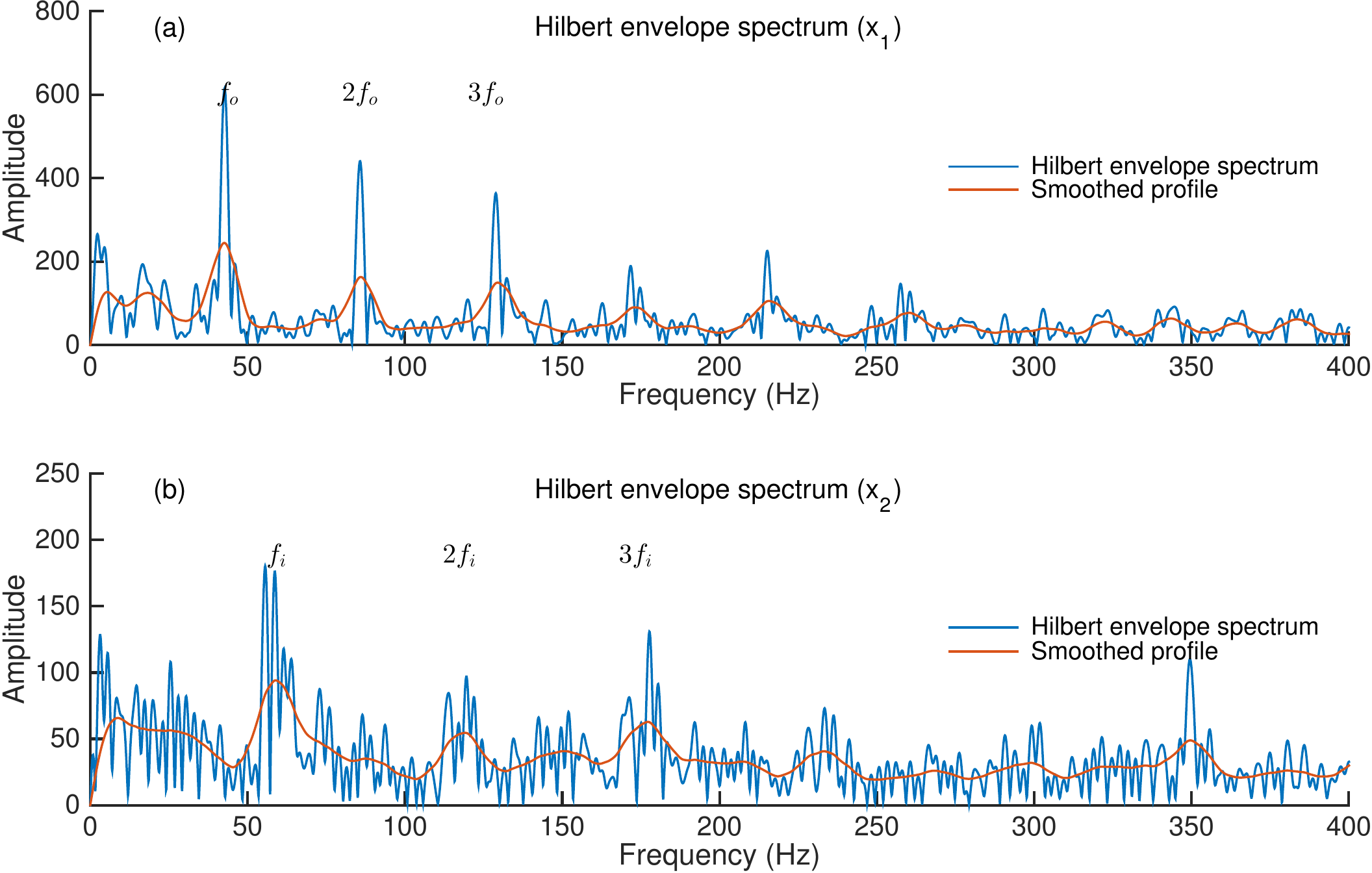}
    \caption{Example~2: Hilbert envelope spectrums of the extracted repetitive transient sequences.}
	\label{fig:fdmca_example_2_hilbert}
\end{figure}

To further reveal the characteristic frequencies,
the Hilbert envelope spectrums of the extracted components are shown in \figref{fig:fdmca_example_2_hilbert}.
The smoothed (by lowpass filtering) profiles of the Hilbert envelope spectrum is also presented to indicate the characteristic frequencies more clearly.
\figref{fig:fdmca_example_2_hilbert}
shows that the characteristic frequencies of outer and inner race are at about 45 and 60 Hz,
and their harmonics are revealed by the peaks of the smoothed profiles.
Thus, the fault features of the two defects are clearly extracted by the proposed approach.

\begin{figure}[t!]
	\centering
    \includegraphics [scale = \figurescale] {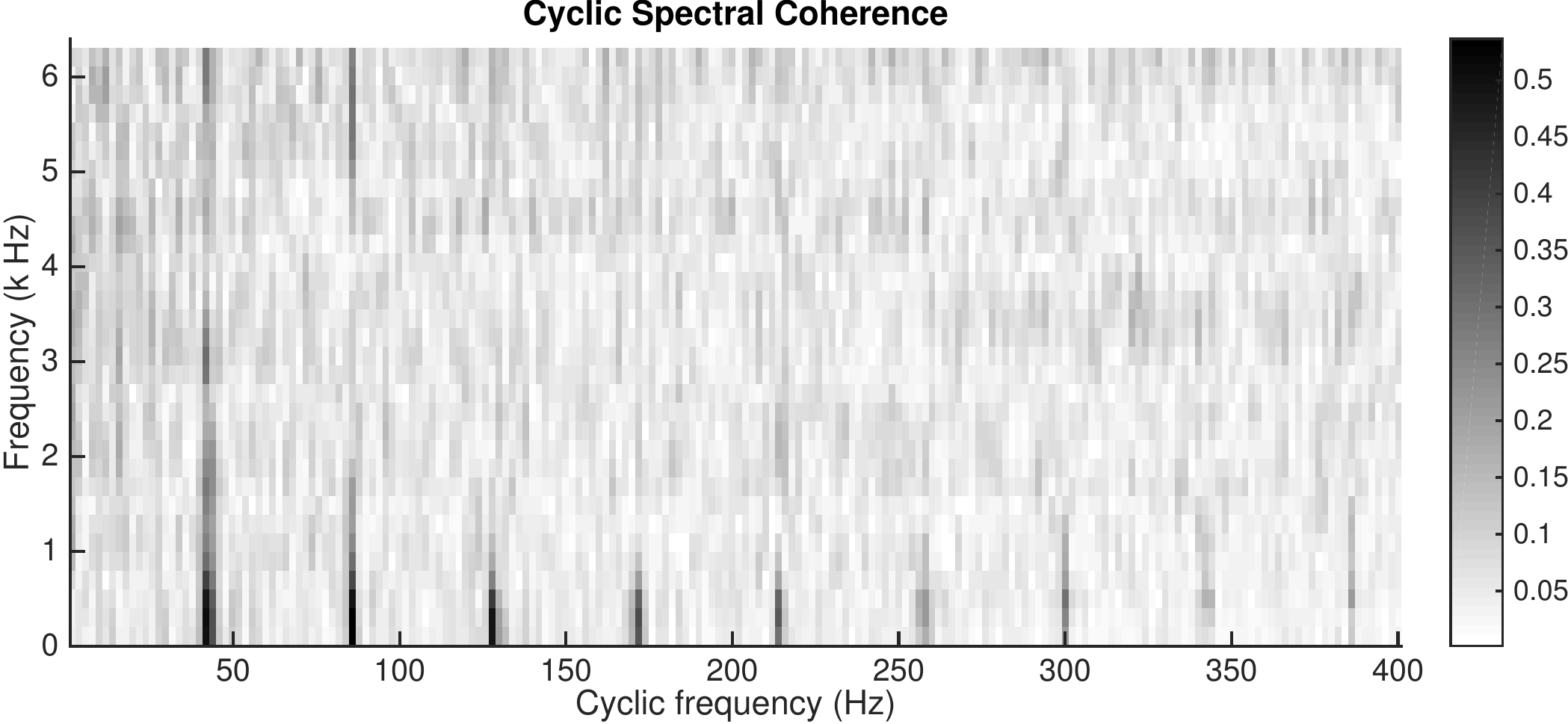}
    \caption{ Example~2: Cyclic spectral coherence of the test signal given in \figref{fig:fdmca_example_2_result}. }
	\label{fig:fdmca_example_2_coh}
\end{figure}

As a comparison, we include the Welch's estimate of the cyclic spectral coherence for the analysis of cyclostationary signals
\cite{fd_Antoni_mssp_2007, fd_Antoni_jsv_2007, fd_Antoni_mssp_2009}.
The result is illustrated in \figref{fig:fdmca_example_2_coh}.
Note that, the signal length is 6400, the window length is 32, the signal is divided into 14 overlapping blocks and the block overlap is 21 with a Hanning window, 
which is suggested in Ref.~\cite{fd_Antoni_mssp_2007}.
When observing the coherence, the cyclic frequency of the outer race (approximate 43.3 Hz) and its harmonic components can be identified.
However, the characteristic frequencies of inner race 58.7 Hz and its harmonic components cannot be observed from \figref{fig:fdmca_example_2_coh}.
This may be caused by the fact that the component of inner race in this case is weak.


\subsection{Example 3: ~Single fault detection}

\begin{figure}[t!]
	\centering
    \includegraphics [scale = \figurescale] {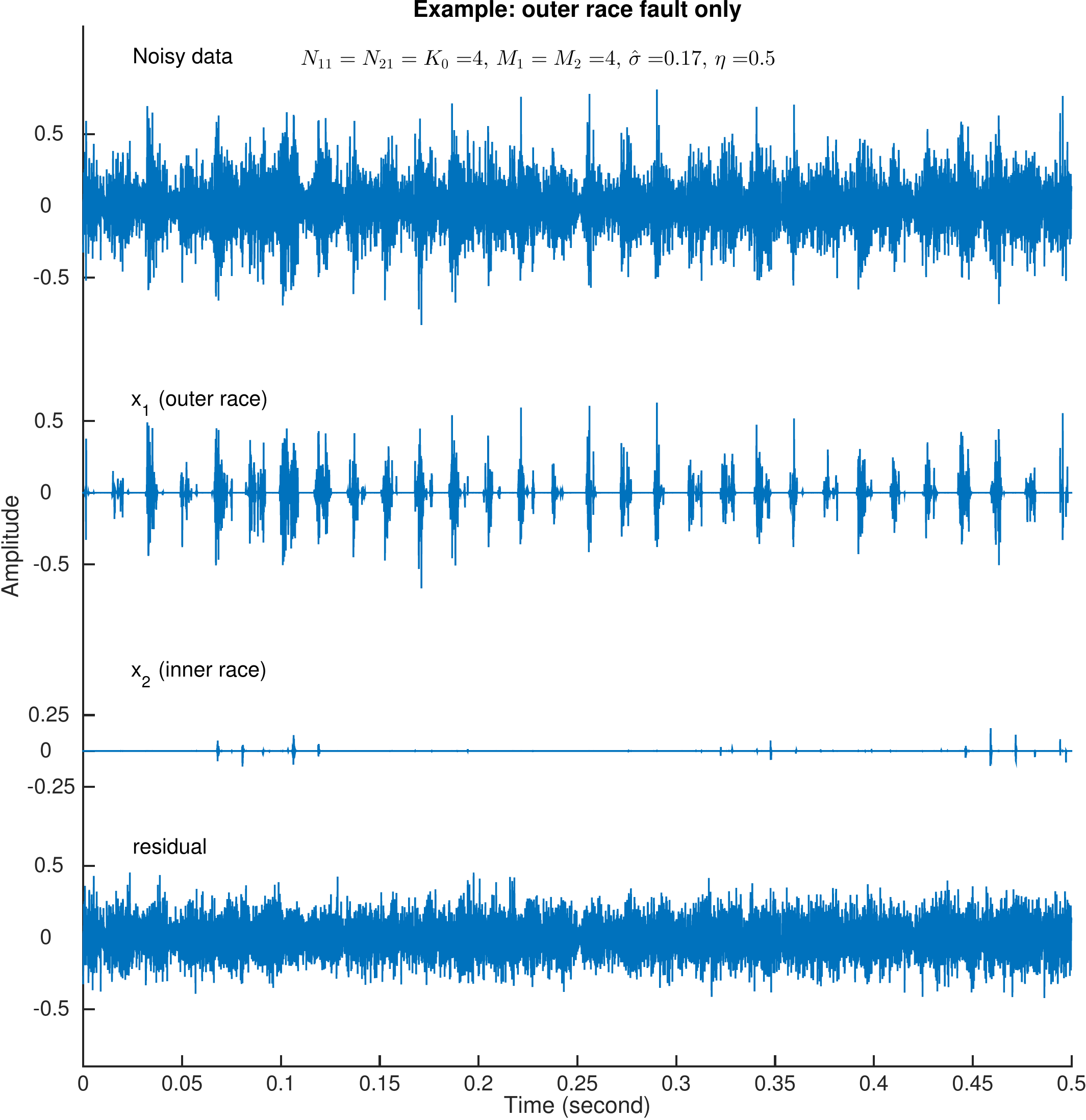}
    \caption{Example~3: Output of RTEA (proposed method) from data with outer race fault only.
    Note that the Y-axis has a different scale to the previous example.}
	\label{fig:fdmca_example_3_result}
\end{figure}

In this example, the proposed method also works when the bearing has only one fault.
\figref{fig:fdmca_example_3_result} shows the measured data and the results from the bearing
in \figref{fig:bearing_fault}, but there was only one defect on the outer race.
The acceleration signals were collected at a constant shaft speed of 481~r/min.
Thus,
the characteristic fault frequencies of the outer and inner races are about $57.8$ Hz and $78.4$ Hz respectively.

Using the given information of the periods,
the repetitive transient sequences $x_1$, $x_2$ and the residual noisy signals are extracted simultaneously from the measured vibration data,
where $x_1$ is corresponds to outer race defect.
Repetitive transients can be observed from the extracted $x_1$,
where the fault frequency can be directly observed around 58 Hz,
because there are about 29 sparse transients evenly distributed with the 0.5 second.
The extracted characteristic frequency 58 Hz is approximately in accordance with the outer race fault frequency of 57.8 Hz.

Moreover, component $x_2$ in \figref{fig:fdmca_example_3_result} is almost purely zero and it exhibits almost no repetitive transients.
This implies that the status of the inner race is healthy.

\section{Conclusion}
This paper proposes a novel approach for the extraction of repetitive transients with group-sparse structure 
in vibration signals, and for detecting faults in rolling element bearings.
%
%
To simultaneously extract both sparse components and perform denoising,
a repetitive group-sparsity based optimization problem is formulated.
To solve the problem,
a computationally efficient iterative algorithm, termed repetitive transients extraction algorithm(RTEA)
is derived.
The non-convex penalty function is used to strongly promote sparsity,
and a condition is given so that the objective function is strictly convex.
Moreover, for practical problems,
an approach to select regularization and non-convexity parameters is provided.

\section*{Appendices}

\appendix
\numberwithin{equation}{section}
\section{Proof of Proposition~\ref{prop:fdmca_convex}}\label{app:fdmca_convex}

\begin{proof}
The first two terms of the objective function in \eqnref{eqn:fdmca_problem} can be rewritten as
\begin{align}\label{eqn:fdmca_ogs_part}
	F(u) = \half \norm{y-u}_2^2
	+ \lam_0 \sum_{n} \phie \Big( \Big[ \sum_{k=0}^{K_0-1} [u]_{n+k}^2 \Big]^{1/2}; a_0 \Big)
\end{align}
where $ u \in \real^{N} $ and $ u - (x_1 +x_2 ) = \mathbf{0}$.
Note that $F(u)$ is exactly identical to GSSD problem
\{see Equation~(20) in \cite[Theorem~1]{Chen_Selesnick_2014_GSSD}\}.
Adopting Theorem~1 and Corollary~2 in \cite{Chen_Selesnick_2014_GSSD},
it can be shown that: when
\begin{equation}
	0 \le a_0 < \frac{1}{K_0\lam_0},
\end{equation}
$F(u)$ is strictly convex.
Moreover, it is immediate that when $a=0$, function \eqnref{eqn:fdmca_Phi} is convex.
As a consequence, an equivalent problem of \eqnref{eqn:fdmca_problem} is considered,
namely
\begin{align}\label{eqn:fdmca_equivalent_prob}
	 \{ u\opt , x_1\opt , x_2\opt \}
	 	& =  \arg \min_{u, x_1 , x_2} F(u) + \sum_{i\in\{ 1,2\}}\lam_i \Phi(x_i; b_i)
	  	\nonumber \\
		&\qquad \text{ such that } u - x_1 -x_2 = \mathbf{0},
\end{align}
which satisfies the convexity condition of equality constrained problem in \cite{Boyd_convex},
and which implies that:
when \eqnref{eqn:fdmca_problem} satisfies \eqnref{eqn:fdmca_convex}, it is a convex problem.
\end{proof}


\section{Derivation of majorizing function $R$ \eqnref{eqn:fdmca_R}} \label{app:fdmca_R_func}

Since $\phie\major$ in \eqnref{eqn:fdmca_phie_major}
majorizes $\phie$, an upper bound of function $R$ can be found as
\begin{align}\label{eqn:fdmca_bar_R}
	\bar R( x_1 , x_2, v ; a_0)
	&= \sum_{n} \phie\major \Big( \Big[ \sum_{k} [x_1+x_2]_{n+k}^2 \Big]^{1/2}, v_n; a_0 \Big)  \nonumber \\
	&= \sum_{n} \Bigg\{ \frac{ 1 }{ 2 \psi(v_n;a_0) } \sum_{k} [x_1 + x_2]_{n+k}^2 \Bigg\}+ C
\end{align}
where $C$ is a constant does not depend on $x_1$ or $x_2$, and for any $v \in \real^{N} $,
\begin{align}\label{eqn:fdmca_inequal_1}
	R( x_1 , x_2 ; a_0)  &\le \bar R( x_1 , x_2, v ; a_0).
\end{align}
After algebraic manipulations, $\bar R$ can be further expressed as
\begin{align}\label{eqn:fdmca_bar_R_2}
	\bar R( x_1 , x_2, v(z) ; a_0) = \half \sum_n r_0(n, z) [x_1+x_2]_n^2 + C,
\end{align}
where vector $v$ is dependent on another vector $z \in \real^{N}$,as
\begin{align}
	[v(z)]_n = \Big[  \sum_{k=0}^{K_0-1} [z]_{n+k}^2\Big]^{1/2},
\end{align}
and $r_0 : \mathbb{Z} \times \real^{N} \to \real, $ is defined by
\begin{align}\label{eqn:fdmca_r0}
	r_0( n, z ) :	 = \sum_{j=0}^{K_0-1}\frac{1}{\psi \Big( [v(z)]_{n-j} ; a_0 \Big) }
					 = \sum_{j=0}^{K_0-1}\frac{1}{\psi \bigg( \Big[  \sum_k [z]_{n-j+k}^2\Big]^{1/2} ; a_0 \bigg) }
\end{align}
which is similar to Equation~(37) of \cite{Chen_Selesnick_2014_GSSD}.

Note that, in this case the inequality of \eqnref{eqn:fdmca_inequal_1} is still valid and the equality holds when
\begin{align}
	z = x_1 +x_2.
\end{align}

Moreover, considering a simple inequality that: for any $\alpha_1, \alpha_2, \beta_1, \beta_2 \in \real$,
\begin{align}\label{eqn:fdmca_inequality}
	(\alpha_1 + \alpha_2)^2 \le (\alpha_1 + \alpha_2)^2 + \big[  (\alpha_1-\beta_1)  -  (\alpha_2-\beta_2)  \big]^2,
\end{align}
whose right side can be re-written as
\begin{align}
	 2\alpha_1^2 +  2\alpha_2^2 - 2(\beta_1 - \beta_2)\alpha_1 - 2(\beta_2 - \beta_1)\alpha_2 + (\beta_1 - \beta_2)^2,
\end{align}
where $\alpha_1$ and $\alpha_2$ are de-coupled and the equality holds when $\alpha_1 = \beta_1$ and $\alpha_2 = \beta_2$.
Then, using the inequality \eqnref{eqn:fdmca_inequality} element-wise,  an upper bound of $\bar R$ can be found as 
\begin{align}\label{eqn:fdmca_R_major_2}
	 	 & R\major ( x_1, x_2, z_1, z_2 ; a_0 ) 		\nonumber  \\			
	= 	& \sum_n \Big\{ r_0(n, z_1+z_2)([x_1]_n^2 + [x_2]_n^2 	
		 - [z_1-z_2]_n [x_1]_n - [z_2-z_1]_n [x_2]_n ) \Big\} + C(z_1,z_2),
\end{align}
where $C(z_1,z_2)$ is a constant only dependent on $z_1$ and $z_2$.
Note that since \eqnref{eqn:fdmca_R_major_2} is an upper bound of $\bar R$, consequently
\begin{align}
	R\major ( x_1, x_2, z_1, z_2 ; a_0 ) \ge R(x_1,x_2 ;a_0).
\end{align}
Furthermore, when $z_1 = x_1$ and $z_2 = x_2$,
using \eqnref{eqn:fdmca_inequality} it can be seen that
$R\major ( x_1, x_2, z_1, z_2 ; a_0 ) = \bar R( x_1 , x_2, v(z_1+z_2); a_0) $,
and this implies that
\begin{align}
	R\major ( x_1, x_2, x_1, x_2 ; a_0 ) = R(x_1,x_2 ;a_0).
\end{align}
Therefore, equation \eqnref{eqn:fdmca_R_major_2} is a majorizer of function $R$ in \eqnref{eqn:fdmca_R}.

\bibliographystyle{plain}

\begin{thebibliography}{10}

\bibitem{abbasion2007rolling}
S.~Abbasion, A.~Rafsanjani, A.~Farshidianfar, and N.~Irani.
\newblock Rolling element bearings multi-fault classification based on the
  wavelet denoising and support vector machine.
\newblock {\em Mech. Syst. Signal Process.}, 21(7):2933--2945, 2007.

\bibitem{fd_Antoni_mssp_2007}
J.~Antoni.
\newblock Cyclic spectral analysis in practice.
\newblock {\em Mech. Syst. Signal Process.}, 21(2):597--630, 2007.

\bibitem{fd_Antoni_jsv_2007}
J.~Antoni.
\newblock Cyclic spectral analysis of rolling-element bearing signals: Facts
  and fictions.
\newblock 304(3–5):497--529, 2007.

\bibitem{fd_Antoni_mssp_2009}
J.~Antoni.
\newblock Cyclostationarity by examples.
\newblock {\em Mech. Syst. Signal Process.}, 23(4):987--1036, 2009.

\bibitem{antoni2006spectral}
J. Antoni and R. B.~Randall.
\newblock The spectral kurtosis: application to the vibratory surveillance and
  diagnostics of rotating machines.
\newblock {\em Mech. Syst. Signal Process.}, 20(2):308--331, 2006.

\bibitem{Bayram_ncvx_icassp_2014}
I.~Bayram, P.-Y. Chen, and I.~W. Selesnick.
\newblock Fused lasso with a non-convex sparsity inducing penalty.
\newblock In {\em Proc.\ ICASSP 2014}, pages 4156--4160, May 2014.

\bibitem{bin2012early}
G.~Bin, J.~Gao, X.~Li, and B. S.~Dhillon.
\newblock Early fault diagnosis of rotating machinery based on wavelet
  packets -empirical mode decomposition feature extraction and neural network.
\newblock {\em Mech. Syst. Signal Process.}, 27:696--711, 2012.

\bibitem{boukra2013statistical}
T. Boukra, A. Lebaroud, and G. Clerc.
\newblock Statistical and neural-network approaches for the classification of
  induction machine faults using the ambiguity plane representation.
\newblock {\em IEEE Trans. Ind. Electron.}, 60(9):4034--4042, 2013.

\bibitem{boustany2005subspace}
R. Boustany and J. Antoni.
\newblock A subspace method for the blind extraction of a cyclostationary
  source: Application to rolling element bearing diagnostics.
\newblock {\em Mech. Syst. Signal Process.}, 19(6):1245--1259, 2005.

\bibitem{boustany2008blind}
R. Boustany and J. Antoni.
\newblock Blind extraction of a cyclostationary signal using reduced-rank
  cyclic regression—a unifying approach.
\newblock {\em Mech. Syst. Signal Process.}, 22(3):520--541, 2008.

\bibitem{Boyd_convex}
S.~Boyd and L.~Vandenberghe.
\newblock {\em Convex Optimization}.
\newblock Cambridge University Press, 2004.

\bibitem{chen2012fault}
B. Chen, Z. Zhang, C. Sun, B. Li, Y. Zi, and Z. He.
\newblock Fault feature extraction of gearbox by using overcomplete rational
  dilation discrete wavelet transform on signals measured from vibration
  sensors.
\newblock {\em Mech. Syst. Signal Process.}, 33:275--298, 2012.

\bibitem{Chen_Selesnick_2014_GSSD}
P.-Y. Chen and I.~W. Selesnick.
\newblock Group-sparse signal denoising: Non-convex regularization, convex
  optimization.
\newblock {\em IEEE Trans.\ Signal Process.}, 62(13):3464--3478, July 2014.

\bibitem{Chen_1994_BP}
S.~Chen and D.~L. Donoho.
\newblock Basis pursuit.
\newblock In {\em 1994 Conference Record of the Twenty-Eighth Asilomar,
  Conference on Signals, Systems and Computers, 1994.}, volume~1, pages 41--44,
  October 1994.

\bibitem{Coifman_1995}
R.~R. Coifman and D.~L. Donoho.
\newblock Translation-invariant de-noising.
\newblock In {\em Wavelet and statistics}, pages 125--150. Springer-Verlag,
  1995.

\bibitem{fd_Cui_jsv_2014}
L.~Cui, J.~Wang, and S.~Lee.
\newblock Matching pursuit of an adaptive impulse dictionary for bearing fault
  diagnosis.
\newblock {\em J. Sound Vib.}, 333(10):2840--2862, 2014.

\bibitem{cui2016quantitative}
L. Cui, N.~Wu, C. Ma, and H. Wang.
\newblock Quantitative fault analysis of roller bearings based on a novel
  matching pursuit method with a new step-impulse dictionary.
\newblock {\em Mech. Syst. Signal Process.}, 68:34--43, 2016.

\bibitem{cui2016vibration}
L. Cui, Y.~Zhang, F. Zhang, J. Zhang, and S. Lee.
\newblock Vibration response mechanism of faulty outer race rolling element
  bearings for quantitative analysis.
\newblock {\em J. Sound Vib.}, 364:67--76, 2016.

\bibitem{Ding_2015_SP}
Y.~Ding and I.~W. Selesnick.
\newblock Sparsity-based correction of exponential artifacts.
\newblock {\em Signal Process.}, 120:236--248, March 2016.

\bibitem{wav_Donoho_93_ideal}
D.~Donoho, I.~Johnstone, and I.~M. Johnstone.
\newblock Ideal spatial adaptation by wavelet shrinkage.
\newblock {\em Biometrika}, 81:425--455, 1993.

\bibitem{feng2013recent}
Z. Feng, M. Liang, and F. Chu.
\newblock Recent advances in time--frequency analysis methods for machinery
  fault diagnosis: A review with application examples.
\newblock {\em Mech. Syst. Signal Process.}, 38(1):165--205, 2013.

\bibitem{FBDN_2007_TIP}
M.~Figueiredo, J.~Bioucas-Dias, and R.~Nowak.
\newblock Majorization-minimization algorithms for wavelet-based image
  restoration.
\newblock {\em IEEE Trans.\ Image Process.}, 16(12):2980--2991, December 2007.

\bibitem{he2013time}
Q. He and X. Wang.
\newblock Time--frequency manifold correlation matching for periodic fault
  identification in rotating machines.
\newblock {\em J. Sound Vib.}, 332(10):2611--2626, 2013.

\bibitem{He_mssp_2016}
W.~He, Y.~Ding, Y.~Zi, and I.~W. Selesnick.
\newblock Sparsity-based algorithm for detecting faults in rotating machines.
\newblock {\em Mech. Syst. Signal Process.}, 72–-73:46--64, May 2016.

\bibitem{he2013tunable}
W. He, Y. Zi, B. Chen, S. Wang, and Z. He.
\newblock {Tunable Q-factor wavelet transform denoising with neighboring
  coefficients and its application to rotating machinery fault diagnosis}.
\newblock {\em Sci. China Technol. Sci.}, 56(8):1956--1965, 2013.

\bibitem{he2014automatic}
W. He, Y. Zi, B. Chen, F. Wu, and Z. He.
\newblock Automatic fault feature extraction of mechanical anomaly on induction
  motor bearing using ensemble super-wavelet transform.
\newblock {\em Mech. Syst. Signal Process.}, 54-55:457--480, 2015.

\bibitem{jiang2015study}
H. Jiang, J. Chen, G. Dong, T. Liu, and G. Chen.
\newblock {Study on Hankel matrix-based SVD and its application in rolling
  element bearing fault diagnosis}.
\newblock {\em Mech. Syst. Signal Process.}, 52:338--359, 2015.

\bibitem{mm_Lange_2000}
K.~Lange, D.~Hunter, and I.~Yang.
\newblock Optimization transfer using surrogate objective functions.
\newblock {\em J. of Comp. Graph. Statist.}, 9:1--20, 2000.

\bibitem{lei2008new}
Y.~Lei, Z.~He, Y.~Zi, and X.~Chen.
\newblock New clustering algorithm-based fault diagnosis using compensation
  distance evaluation technique.
\newblock {\em Mech. Syst. Signal Process.}, 22(2):419--435, 2008.

\bibitem{lei2013review}
Y. Lei, J. Lin, Z. He, and M.~J Zuo.
\newblock A review on empirical mode decomposition in fault diagnosis of
  rotating machinery.
\newblock {\em Mech. Syst. Signal Process.}, 35(1):108--126, 2013.

\bibitem{li2011virtual}
Z. Li, X. Yan, C. Yuan, Z. Peng, and L.~Li.
\newblock Virtual prototype and experimental research on gear multi-fault
  diagnosis using wavelet-autoregressive model and principal component analysis
  method.
\newblock {\em Mech. Syst. Signal Process.}, 25(7):2589--2607, 2011.

\bibitem{fd_Liang_mssp_2010}
M.~Liang and I.~Soltani Bozchalooi.
\newblock An energy operator approach to joint application of amplitude and
  frequency-demodulations for bearing fault detection.
\newblock {\em Mech. Syst. Signal Process.}, 24(5):1473 -- 1494, 2010.

\bibitem{malhi2004pca}
A. Malhi and R.~X. Gao.
\newblock {PCA-based feature selection scheme for machine defect
  classification}.
\newblock {\em IEEE Trans. Instrum. Meas.}, 53(6):1517--1525, 2004.

\bibitem{fd_Qin_jsv_2013}
Y.~Qin, Y.~Mao, and B.~Tang.
\newblock Vibration signal component separation by iteratively using basis
  pursuit and its application in mechanical fault detection.
\newblock {\em J. Sound Vib.}, 332(20):5217--5235, 2013.

\bibitem{randall2011rolling}
R.~B. Randall and J. Antoni.
\newblock Rolling element bearing diagnostics -- a tutorial.
\newblock {\em Mech. Syst. Signal Process.}, 25(2):485--520, 2011.

\bibitem{selesnick2014sparse}
I. W. Selesnick and I. Bayram.
\newblock Sparse signal estimation by maximally sparse convex optimization.
\newblock {\em IEEE Trans.\ Signal Process.}, 62(5):1078--1092, 2014.

\bibitem{selesnick2015convex}
I.~W. Selesnick, A. Parekh, and I. Bayram.
\newblock Convex 1-d total variation denoising with non-convex regularization.
\newblock {\em IEEE Signal Processing Letters}, 22(2):141--144, 2015.

\bibitem{smith2015rolling}
W.~A. Smith and R.~B. Randall.
\newblock Rolling element bearing diagnostics using the case western reserve
  university data: A benchmark study.
\newblock {\em Mech. Syst. Signal Process.}, 64:100--131, 2015.

\bibitem{mca_Starck_2004}
J.-L. Starck, M.~Elad, and D.~Donoho.
\newblock Redundant multiscale transforms and their application for
  morphological component analysis.
\newblock {\em Advances in Imaging and Electron Physics}, 132:287--348, 2004.

\bibitem{fd_Su_mssp_2010}
W.~Su, F.~Wang, H.~Zhu, Z.~Zhang, and Z.~Guo.
\newblock Rolling element bearing faults diagnosis based on optimal morlet
  wavelet filter and autocorrelation enhancement.
\newblock {\em Mech. Syst. Signal Process.}, 24(5):1458--1472, 2010.

\bibitem{Tibshirani_2005}
R.~Tibshirani, M.~Saunders, S.~Rosset, J.~Zhu, and K.~Knight.
\newblock Sparsity and smoothness via the fused lasso.
\newblock {\em Journal of the Royal Statistical Society Series B}, pages
  91--108, 2005.

\bibitem{wang2011constrained}
Z. Wang, J. Chen, G. Dong, and Y.~Zhou.
\newblock Constrained independent component analysis and its application to
  machine fault diagnosis.
\newblock {\em Mech. Syst. Signal Process.}, 25(7):2501--2512, 2011.

\bibitem{yan2010harmonic}
R. Yan and R.~X. Gao.
\newblock Harmonic wavelet-based data filtering for enhanced machine defect
  identification.
\newblock {\em J. Sound Vib.}, 329(15):3203--3217, 2010.

\bibitem{yan2014wavelets}
R. Yan, R.~X. Gao, and X. Chen.
\newblock Wavelets for fault diagnosis of rotary machines: A review with
  applications.
\newblock {\em Signal Process.}, 96:1--15, 2014.

\bibitem{fd_Yang_mssp_2005}
H.~Yang, J.~Mathew, and L.~Ma.
\newblock Fault diagnosis of rolling element bearings using basis pursuit.
\newblock {\em Mech. Syst. Signal Process.}, 19(2):341--356, 2005.

\bibitem{zhang2013multi}
X. Zhang and J. Zhou.
\newblock Multi-fault diagnosis for rolling element bearings based on ensemble
  empirical mode decomposition and optimized support vector machines.
\newblock {\em Mech. Syst. Signal Process.}, 41(1):127--140, 2013.

\end{thebibliography}

\end{document}